\documentclass[12pt,final,floatfix,eqsecnum,onecolumn,amsfonts,amsmath,amssymb]{revtex4}

\usepackage[sf,bf,small]{titlesec}
\usepackage{mathrsfs}
\usepackage{bm}
\usepackage{bbm}
\usepackage{xr}

\newcommand{\extref}[1]{II.\ref{#1}} 

\providecommand*{\I}{\mathrm{i}}                           
\providecommand*{\rbra}[1]{(#1|}                           
\providecommand*{\rket}[1]{|#1)}                           
\providecommand*{\lrbra}[1]{\rbra{\widetilde{#1}}}    
\providecommand*{\rbracket}[2]{\rbra{#1}#2)}          
\providecommand*{\lrbracket}[2]{\lrbra{#1}#2)}       

\providecommand*{\klr}[1]{\left(#1\right)}					 
\providecommand*{\kle}[1]{\left[#1\right]}					 
\providecommand*{\klg}[1]{\left\{#1\right\}}				 

\providecommand*{\mrmd}{d}									 
\providecommand*{\euler}{\mathrm{e}}								 

\DeclareMathOperator{\mIm}{Im}												
\DeclareMathOperator{\Tr}{Tr}													
\DeclareMathOperator{\diag}{diag}											

\newcommand*{\mcal}[1]{\mathcal{#1}}                   

\newcommand*{\umat}[1]{\underline{\mathscr{#1}}}      
\renewcommand{\vec}[1]{\bm{#1}}                       
\newcommand*{\uvec}[1]{\underline{\bm{#1}}}           

\renewcommand{\thesection}{\arabic{section}}

\titlespacing{\section}{0pt}{40pt}{20pt}

\begin{document}

\title{\Large Metastable states, the adiabatic theorem\\ and parity violating geometric phases I}

\author{Timo~Bergmann}
\email{T.Bergmann@ThPhys.Uni-Heidelberg.DE}

\author{Thomas~Gasenzer}
\email{T.Gasenzer@ThPhys.Uni-Heidelberg.DE}

\author{Otto~Nachtmann}
\email{O.Nachtmann@ThPhys.Uni-Heidelberg.DE}

\affiliation{Institut f{\"u}r Theoretische Physik, Universit{\"a}t Heidelberg,\\ 
Philosophenweg 16, 69120 Heidelberg, Germany}

\date{\today}

\begin{abstract}
A system of metastable plus unstable states is discussed. The mass matrix governing the time development of the system is supposed to vary slowly with time. The adiabatic limit for this case is studied and it is shown that only the metastable states obtain the analogs of the dynamical and geometrical phase factors familiar from stable states. Abelian and non-abelian geometric phase factors for metastable states are defined.\\[-7ex]

\hfill
{\small HD--THEP--07--08}
\end{abstract}

\maketitle

\section{Introduction}\label{s:Introduction}

The investigations reported in this paper are in connection with an effort to measure atomic parity-(P-)violating effects with an atomic beam interferometer described in \cite{ABSE95}. Parity violation in atoms through neutral current exchange was already studied in \cite{Zel59}. The seminal paper in this field is \cite{Bou74}, for reviews see \cite{Khrip91,Bou97,Bou05}. Measurements of P-violating effects in heavy many electron atoms like Cs and Tl have already reached a high precision \cite{BeWi99,Bou05,Vet95}. For atoms with a single electron like hydrogen or deuterium, however, experimental measurements have, so far, not been successful. In an atomic beam apparatus an atom can be subjected to external electric and magnetic fields. Under suitable conditions the motion of the atom can be described in the adiabatic limit. As is well known from Berry's work \cite{Ber84} the atomic wave function can then acquire a geometric phase factor. For a collection of important papers on geometric phases see \cite{ShWi89}. We shall study how such phase factors occur for metastable states and classify the phases in parity conserving (PC) and parity violating (PV) ones. The final aim is to see if PV geometric phases are large enough to be measurable with an atomic beam apparatus. But this will be dealt with in future work. 

Our paper is organised as follows. In section \ref{s:AdiabaticTheorem} we discuss the description of unstable atoms in slowly varying external electric and magnetic fields. We discuss the adiabatic limit for stable states and a set of unstable states all having the same decay rate in section 3. In section 4 we deal with a simple example of a two state system where  one member is longer lived than the other one. The more general case of a number of metastable states with equal decay rates and a number of unstable states with larger decay rates is treated in section 5. Our conclusions are presented in section 6. Two appendices present the detailed mathematical arguments leading to the results of sections 4 and 5. In the companion paper II \cite{BeGaNa07_II} we study the metastable $2S$ states of hydrogen and deuterium and identify the PC and PV phases occurring there. In the following we shall refer to tables of II by table II.1 etc., to equations as (II.1.1) etc.

We use units $\hbar=c=1$ if other units are not explicitly indicated.

\section{Metastable states, Generalities}\label{s:AdiabaticTheorem}\label{s:Generalities}

Let us consider an atomic system with $N$ stable or unstable states. We assume that the atom is at rest and subjected to slowly varying external electric and magnetic fields. The effective Schr\"odinger equation for the system is then, in the Wigner-Weisskopf approximation,
\begin{align}\label{e2:WW.SE}
\I\frac{\partial}{\partial t}\rket{t} = \umat M(t)\rket{t}\ ,
\end{align}
where $\rket{t}$ is the state vector of the undecayed states and $\umat M(t)$ the mass (or complex energy) matrix which is, in general, non hermitian. For a discussion of the complete Wigner-Weisskopf solution see for instance \cite{BoBrNa95}. 

To give an example which we shall study in detail in II let us consider the states with principal quantum number $n=2$ in ordinary hydrogen. As basis states we have here the $2S$ and $2P$ states. In total these are $N=16$ states which we number as shown in table \extref{II.t:state.labels}. 
In table \extref{II.t:H2.M0} 
we also give the mass matrix $\umat M_0$ for the case of zero external fields. With external fields the mass matrix is
\begin{align}\label{e2:MM}
\umat M(t)=\umat M_0-\uvec D\cdot \vec{\mcal E}(t) - \uvec\mu \cdot\vec{\mcal B}(t)\ .
\end{align}
Here $\vec{{\mcal E}}(t)$ is the electric, $\vec{\mcal B}(t)$ the magnetic field strength vector. These are supposed to vary only slowly with $t$. Furthermore $\uvec D$ is the matrix of the electric and $\uvec\mu$ of the magnetic dipole operators in the space of $n=2$ states, see tables \extref{II.t:H2.D} 
and \extref{II.t:H2.Mu}, 
respectively.

We return to the general case (\ref{e2:WW.SE}). We want to study the adiabatic limit, that is, a given change of $\umat M(t)$ with $t$ is made over a longer and longer time. Following \cite{Messiah2} we implement this by introducing a reduced time $\tau$, setting 
\begin{align}\label{e2:red.time}
t=\frac{T}{\tau_0}\tau
\end{align}
where $\tau_0$ is some fixed time. We consider the system over the interval
\begin{align}\label{e2:tau.int}
0\leq\tau\leq\tau_0
\end{align}
in the limit of larger and larger $T$. The mass matrix in (\ref{e2:WW.SE}) is supposed to be a function of $\tau$ only 
\begin{align}\label{e2:M.tau}
\umat M(t)=\umat{\hat M}(\tau)\ .
\end{align}
We get then from (\ref{e2:WW.SE}) with $\rket{t}\equiv\rket{T;\tau}$
\begin{align}\label{e2:SE.tau}
\I\frac{\partial}{\partial\tau}\rket{T;\tau}=\frac{T}{\tau_0}\umat{\hat M}(\tau)\rket{T;\tau}\ .
\end{align}
Here and in the following we indicate explicitly the dependence of the quantities on $\tau$ and $T$, respectively.

The mass matrix $\umat{\hat M}(\tau)$ (\ref{e2:M.tau}) can be decomposed into the hermitian energy and decay matrices
\begin{align}\label{e2:MM.decomp}
\begin{split}
\umat{\hat M}(\tau) &= \underline{E}(\tau)-\frac{\I}{2}\underline{\Gamma}(\tau)\ ,\\
\underline{E}^\dagger(\tau) &= \underline{E}(\tau)\ ,\\
\underline{\Gamma}^\dagger(\tau) &= \underline{\Gamma}(\tau)\ .
\end{split}
\end{align}
We suppose that $\underline{\Gamma}(\tau)$  is a positive-semidefinite matrix
\begin{align}\label{e2:Gamma.psd}
\underline{\Gamma}(\tau)=\I\big(\umat{\hat M}(\tau)-\umat{\hat M}^\dagger(\tau)\big)\geq 0\ .
\end{align}
We suppose, furthermore, that for all $\tau$ $\umat{\hat M}(\tau)$ can be diagonalised. (This is, for instance, guaranteed if $\umat{\hat M}(\tau)$ has $N$ different eigenvalues.) Then we have for all $\tau$ complete sets of right and left eigenvectors satisfying
\begin{align}\label{e2:LEV}
\umat{\hat M}(\tau)\rket{\alpha,\tau}  &= E(\alpha,\tau)\rket{\alpha,\tau}\ ,\\ \label{e2:REV}
\lrbra{\alpha,\tau}\umat{\hat M}(\tau) &= \lrbra{\alpha,\tau}E(\alpha,\tau)\ ,\\ \nonumber
(\alpha &= 1,\dots,N)\ .
\end{align}
Here $E(\alpha,\tau)$ are the --- in general complex --- eigenvalues of $\umat{\hat M}(\tau)$,
\begin{align}\label{e2:EW}
E(\alpha,\tau)=E_R(\alpha,\tau)-\frac{\I}{2}\Gamma(\alpha,\tau),
\end{align}
with $E_R(\alpha,\tau)$ the real part of the energy and $\Gamma(\alpha,\tau)$ the decay rate of the state $|\alpha,\tau)$. We choose the normalisation of the eigenvectors such that
\begin{align}\label{e2:LR.ON}
\lrbracket{\alpha,\tau}{\beta,\tau} &= \delta_{\alpha\beta}\ ,\\ \label{e2:RR.Norm}
\rbracket{\alpha,\tau}{\alpha,\tau} &= 1\quad\qquad
\text{(no summation over $\alpha$)\ .}
\end{align}
Everything is supposed to be continuous in $\tau$ and we shall use this in the following in an essential way.

From (\ref{e2:Gamma.psd}) follows 
\begin{align}\label{e2:GEW.pos}
\Gamma(\alpha,\tau)\geq 0\ ,\quad(\alpha=1,\dots,N)\ .
\end{align}
The proof is simple. From (\ref{e2:Gamma.psd})-(\ref{e2:RR.Norm}) we obtain
\begin{align}\label{e2:G.psd.proof}
\begin{split}
\rbra{\alpha,\tau}\underline{\Gamma}(\tau)\rket{\alpha,\tau}
&= \rbra{\alpha,\tau}\I\klr{\umat{\hat M}(\tau)-\umat{\hat M}^\dagger(\tau)}\rket{\alpha,\tau} 
= \Gamma(\alpha,\tau) 
\geq 0\ .
\end{split}
\end{align}
Note that the reverse is in general not true, that is, (\ref{e2:Gamma.psd}) does not follow from (\ref{e2:GEW.pos}). 

Now we expand the state vector $\rket{T;\tau}$ in terms of the eigenstates $\rket{\alpha,\tau}$:
\begin{align}\label{e2:State.Expansion}
\rket{T;\tau} = \sum_{\alpha=1}^N\psi_\alpha(T;\tau)\rket{\alpha,\tau}\ .
\end{align}
From  (\ref{e2:SE.tau}) we get then
\begin{align}\label{e2:SE.final}
\I\frac{\partial}{\partial\tau}\psi_\alpha(T;\tau) = 
\frac{T}{\tau_0}E(\alpha,\tau)\psi_\alpha(T;\tau)
- \sum^N_{\beta=1}\lrbra{\alpha,\tau}\I\frac{\partial}{\partial\tau}\rket{\beta,\tau}\psi_\beta(T;\tau)\ .
\end{align}
The task is to discuss the solution of (\ref{e2:SE.final}) for large $T$. 

\section{Stable states and states with equal decay rates}\label{s:StableStates}

Here we discuss briefly the cases that all states are stable or that they are unstable but have equal decay rates. In detail we suppose the following
\begin{align}\label{e2:GammasEqual}
\Gamma(\alpha,\tau)-\Gamma(\beta,\tau)=0
\end{align}
for all $\alpha,\beta\in\klg{1,\dots,N}$ and all $\tau\in\kle{0,\tau_0}$. This includes the case of stable states where $\Gamma(\alpha,\tau)=0$. Furthermore we suppose
\begin{align}\label{e2:EW.non.deg}
|E(\alpha,\tau)-E(\beta,\tau)|\geq c>0
\end{align}
for all $\alpha\neq\beta$ and all $\tau\in\kle{0,\tau_0}$, where $c$ is a positive constant. With (\ref{e2:GammasEqual}) and (\ref{e2:EW.non.deg}) the discussion of the adiabatic limit of the solution of (\ref{e2:SE.final}) can be done exactly as in \cite{Messiah2} and one finds as solution of (\ref{e2:SE.final}) for large $T$
\begin{align}\label{e2:SE.Solution.Stable}
\begin{split}
\psi_\alpha(T;\tau) &= \exp\kle{-\I\frac{T}{\tau_0}\varphi_\alpha(\tau)
+\I\gamma_{\alpha\alpha}(\tau)}\klg{\psi_\alpha(T;0)+\mathcal O\klr{\frac{1}{T}}}\ ,\\
(\alpha&=1,\dots,N)\ .
\end{split}
\end{align}
Here $T\varphi_\alpha(\tau)/\tau_0$ is the dynamical and $\gamma_{\alpha\alpha}(\tau)$ is the geometric (Berry-)phase,
\begin{align}\label{e2:Dyn.Phase}
\varphi_\alpha(\tau) &= \int^\tau_0\mrmd\tau'~E(\alpha,\tau')\ ,\\ \label{e2:Berry.Phase}
\gamma_{\alpha\alpha}(\tau) &= \int^\tau_0\mrmd\tau'~\lrbra{\alpha,\tau'}\I\frac{\partial}{\partial \tau'}\rket{\alpha,\tau'}\ .
\end{align}
From (\ref{e2:SE.Solution.Stable}) we see that for large $T$ the following holds. If we start with an eigenstate of $\umat{\hat M}(0)$ for $\tau=0$ the system will be in the corresponding eigenstate of $\umat{\hat M}(\tau_0)$ for $\tau=\tau_0$. For decaying states (satisfying (\ref{e2:GammasEqual})) both, the dynamical and the geometric phase factors will have real \emph{and} imaginary parts. That is, also the effective decay rates of the states $\alpha$ will obtain a geometric contribution. The real, but not the imaginary part of the geometric phase is ``gauge'' dependent. If we change the definition of the states $\rket{\alpha,\tau}$ by
\begin{align}\label{e2:Gauge.Trafo}
\begin{split}
\rket{\alpha,\tau} &\quad\longrightarrow\quad \rket{\alpha,\tau}' = \euler^{\I\eta_\alpha(\tau)}\rket{\alpha,\tau}\ ,\\
\lrbra{\alpha,\tau} &\quad\longrightarrow\quad \lrbra{\alpha,\tau}' = \euler^{-\I\eta_\alpha(\tau)}\lrbra{\alpha,\tau}\ ,
\end{split}
\end{align}
where $\eta_\alpha(\tau)$ must be real in order to respect (\ref{e2:RR.Norm}) we get
\begin{align}\label{e2:BP.Int.Trafo}
\lrbra{\alpha,\tau}\I\frac{\partial}{\partial\tau}\rket{\alpha,\tau}
\quad\longrightarrow\quad 
\lrbra{\alpha,\tau}\I\frac{\partial}{\partial\tau}\rket{\alpha,\tau} 
-\frac{\partial\eta_\alpha(\tau)}{\partial\tau}\ .
\end{align}
The change of the geometric phase (\ref{e2:Berry.Phase}) induced by the transformation (\ref{e2:Gauge.Trafo}) of the basis states is, therefore, 
\begin{align}\label{e2:BP.Gauge.Trafo}
\gamma'_{\alpha\alpha}(\tau)=\gamma_{\alpha\alpha}(\tau)-\eta_\alpha(\tau)+\eta_\alpha(0)\ .
\end{align}

\section{A two state system of decaying states}\label{s:TwoStates}

Before going to the general case of $N$ decaying states we discuss as an example a system with two unstable states $(N=2)$. Let the state 1 be the longer-lived one. In detail we suppose
\begin{align}\label{e2:Decay.Assumption}
\Gamma(2,\tau)-\Gamma(1,\tau)\geq\Delta \Gamma_{\mathrm{min}}>0
\end{align}
for all $\tau\in\kle{0,\tau_0}$, where $\Delta\Gamma_{\mathrm{min}}$ is a fixed constant. The real parts of the energies, $E_R(1,\tau)$ and $E_R(2,\tau)$, see (\ref{e2:EW}), can be arbitrary. From (\ref{e2:SE.final}) we get for $N=2$ the coupled equations
\begin{align}\label{e2:Two.States.SE}
\begin{split}
\I\frac{\partial\psi_1}{\partial\tau}(T;\tau) &= \kle{\frac{T}{\tau_0}E(1,\tau)-a_{11}(\tau)}\psi_1(T;\tau)
-a_{12}(\tau)\psi_2(T;\tau)\ ,\\
\I\frac{\partial\psi_2}{\partial\tau}(T;\tau) &= \kle{\frac{T}{\tau_0}E(2,\tau)-a_{22}(\tau)}\psi_2(T;\tau)
-a_{21}(\tau)\psi_1(T;\tau)\ .
\end{split}
\end{align}
Here and in the following we use the definitions
\begin{align}\label{e2:Def.a}
a_{\alpha\beta}(\tau) &= \lrbra{\alpha,\tau}\I\frac{\partial}{\partial\tau}\rket{\beta,\tau}\ ,\\ \label{e2:Def.gamma}
\gamma_{\alpha\beta}(\tau) &= \int^\tau_0\mrmd\tau'~a_{\alpha\beta}(\tau')\ ,\\ \label{e2:Def.varphi}
\begin{split}
\varphi_\alpha(\tau)&= \int^\tau_0\mrmd\tau'~E(\alpha,\tau')\\
&= \int^\tau_0\mrmd\tau'\kle{E_R(\alpha,\tau')-\frac{\I}{2}\Gamma(\alpha,\tau')}\ ,
\end{split}\\ \label{e2:Def.rho}
\rho_\alpha(\tau) &= \int^\tau_0\mrmd\tau'~\Gamma(\alpha,\tau')\ ,
\end{align}
where $\alpha,\beta\in\klg{1,2}$. (In later chapters we use these same definitions for $\alpha,\beta\in\klg{1,\ldots,N}$) We have
\begin{align}\label{e2:Im.varphi}
\mIm\varphi_\alpha(\tau)=-\frac{1}{2}\rho_\alpha(\tau).
\end{align}

With the initial conditions $\psi_\alpha(T;0)$ we can transform (\ref{e2:Two.States.SE}) into integral equations
\begin{align}\label{e2:2St.ISE.1}
\begin{split}
\psi_1(T;\tau) &= \exp\kle{-\I\frac{T}{\tau_0}\varphi_1(\tau)+\I\gamma_{11}(\tau)}\\
&\times\klg{\psi_1(T;0)+\int^\tau_0\mrmd\tau'~\exp\kle{\I\frac{T}{\tau_0}\varphi_1(\tau')-\I\gamma_{11}(\tau')}
\I a_{12}(\tau')\psi_2(T;\tau')}\ ,
\end{split}\\ \label{e2:2St.ISE.2}
\begin{split}
\psi_2(T;\tau) &= \exp\kle{-\I\frac{T}{\tau_0}\varphi_2(\tau)+\I\gamma_{22}(\tau)}\\
&\times \klg{\psi_2(T;0)+\int^\tau_0\mrmd\tau'~\exp\kle{\I\frac{T}{\tau_0}\varphi_2(\tau')-\I\gamma_{22}(\tau')}
\I a_{21}(\tau')\psi_1(T;\tau')}\ .
\end{split}
\end{align}
Inserting (\ref{e2:2St.ISE.2}) in (\ref{e2:2St.ISE.1}) we get an equation for $\psi_1(T;\tau)$ alone. To write this in a transparent form we introduce functions $\chi_1(T;\tau)$ and $\chi^{(0)}_1(T;\tau)$ by 
\begin{align}\label{e2:Def.chi1}
\psi_1(T;\tau) &= \exp\kle{-\I\frac{T}{\tau_0}\varphi_1(\tau)+\I\gamma_{11}(\tau)}\chi_1(T;\tau)\ ,\\ \label{e2:Def.chi1.0}
\chi^{(0)}_1(T;\tau) &= \psi_1(T;0)+l_{12}(T;\tau)\psi_2(T;0)\ ,
\end{align}
where
\begin{align}\label{e2:Def.l12}
l_{12}(T;\tau)=\I\int^\tau_0\mrmd\tau_1~\exp\kle{\I\frac{T}{\tau_0}\big(\varphi_1(\tau_1)-\varphi_2(\tau_1)\big)
-\I\gamma_{11}(\tau_1)+\I\gamma_{22}(\tau_1)}a_{12}(\tau_1)\ .
\end{align}
Furthermore we define an integral operator $L$ for continuous functions $\tau\to\zeta(\tau)$ $(\tau\in\kle{0,\tau_0})$ as follows
\begin{align}\label{e2:Def.L}
\begin{split}
(L\zeta)(\tau) &= -\int^\tau_0\mrmd\tau_1~
\exp\kle{\I\frac{T}{\tau_0}\big(\varphi_1(\tau_1)-\varphi_2(\tau_1)\big)
-\I\gamma_{11}(\tau_1)+\I\gamma_{22}(\tau_1)}a_{12}(\tau_1)\\
&\times \int^{\tau_1}_0\mrmd\tau_2~
\exp\kle{-\I\frac{T}{\tau_0}\big(\varphi_1(\tau_2)-\varphi_2(\tau_2)\big)
+\I\gamma_{11}(\tau_2)-\I\gamma_{22}(\tau_2)}a_{21}(\tau_2)\ \zeta(\tau_2)\ .
\end{split}
\end{align}
We get then from (\ref{e2:2St.ISE.1}) and (\ref{e2:2St.ISE.2}) 
\begin{align}\label{e2:chi.L.chi0}
\chi_1(T;\tau)=\chi^{(0)}_1(T;\tau)+(L\chi_1)(T;\tau)\ .
\end{align}
From (\ref{e2:chi.L.chi0}) we obtain
\begin{align}\label{e2:chi0}
\big((1-L)\chi_1\big)(T;\tau)=\chi^{(0)}_1(T;\tau)
\end{align}
with the solution
\begin{align}\label{e2:chi1.solution}
\chi_1(T;\tau)=\klr{(1-L)^{-1}\chi^{(0)}_1}(T;\tau)
=\chi^{(0)}_1(T,\tau)
+\sum^\infty_{n=1}\klr{L^n\chi^{(0)}_1}(T;\tau)\ .
\end{align}
With (\ref{e2:Def.chi1.0}) this gives
\begin{align}\label{e2:chi1}
\chi_1(T;\tau)=\psi_1(T;0)+l_{12}(T;\tau)\psi_2(T;0)
+\sum^\infty_{n=1}\klr{L^n\chi^{(0)}_1}(T;\tau)\ .
\end{align}

Now we show that all terms on the r.h.s. of (\ref{e2:chi1}) except $\psi_1(T;0)$ are of order $1/T$ for large $T$. We note first that from (\ref{e2:Def.varphi}) and (\ref{e2:Def.rho}) we have
\begin{align}\label{e2:abs.exp}
\left|\exp\kle{\pm\I\frac{T}{\tau_0}\varphi_\alpha(\tau)}\right|
= \exp\kle{\pm\frac{1}{2}\frac{T}{\tau_0}\rho_\alpha(\tau)}\,,\quad (\alpha=1,2)\ .
\end{align}
We get now from (\ref{e2:Def.l12}) and (\ref{e2:Decay.Assumption})
\begin{align}\label{e2:l12.bound}
\begin{split}
\left|l_{12}(T;\tau)\right| &\leq
\int^\tau_0\mrmd\tau_1~\exp\kle{-\frac{1}{2}\frac{T}{\tau_0}\klr{\rho_2(\tau_1)-\rho_1(\tau_1)}}\\
&\quad\times \left|\exp\Big(-\I\gamma_{11}(\tau_1)+\I\gamma_{22}(\tau_1)\Big)\right|\,\left|a_{12}(\tau_1)\right|\\
&\leq \int^\tau_0 \mrmd\tau_1~\exp\kle{-\frac{1}{2}\frac{T}{\tau_0}\klr{\rho_2(\tau_1)-\rho_1(\tau_1)}}\\
&\quad\times\klr{\Gamma(2,\tau_1)-\Gamma(1,\tau_1)}\Delta\Gamma^{-1}_{\mathrm{min}}\\
&\quad\times\left|\exp\Big(-\I\gamma_{11}(\tau_1)+\I\gamma_{22}(\tau_1)\Big)\right|\,\left|a_{12}(\tau_1)\right|\ .
\end{split}
\end{align}
Above, after (\ref{e2:RR.Norm}), we have explicitly supposed that all functions like $a_{12}(\tau),\gamma_{11}(\tau)$ etc. are continuous for $\tau\in\kle{0,\tau_0}$. Thus they are bounded there. We set
\begin{align}\label{e2:Gmin.bound}
c_{12} = \max_{\tau_1\in\kle{0,\tau_0}}
\klg{\tau_0\left|\exp\Big(-\I\gamma_{11}(\tau_1)+\I\gamma_{22}(\tau_1)\Big)\right|\,\left|a_{12}(\tau_1)\right|}\ .
\end{align}
From (\ref{e2:l12.bound}) we get then with (\ref{e2:Def.rho}) 
\begin{align}\label{e2:l12.bound.final}
\begin{split}
\left|l_{12}(T;\tau)\right| &\leq c_{12}\int^\tau_0 \mrmd\tau_1~
\exp\kle{-\frac{T}{2\tau_0}\big(\rho_2(\tau_1)-\rho_1(\tau_1)\big)}
\kle{\Gamma(2,\tau_1)-\Gamma(1,\tau_1)}(\tau_0\Delta\Gamma_{\rm min})^{-1}\\
&= c_{12}\int^\tau_0\mrmd\tau_1~\exp\kle{-\frac{T}{2\tau_0}\big(\rho_2(\tau_1)-\rho_1(\tau_1)\big)}
\frac{\mrmd}{\mrmd\tau_1}\big(\rho_2(\tau_1)-\rho_1(\tau_1)\big) (\tau_0\Delta\Gamma_{\rm min})^{-1}\\
&= c_{12}\frac{2}{T}\klr{1-\exp\kle{-\frac{T}{2\tau_0}\big(\rho_2(\tau)-\rho_1(\tau)\big)}}(\Delta\Gamma_{\rm min})^{-1}\\
&\leq c_{12}\frac{2}{T}(\Delta\Gamma_{\rm min})^{-1}\ .
\end{split}
\end{align}
Here we use that due to (\ref{e2:Decay.Assumption}) and (\ref{e2:Def.rho}) we have for all $\tau\in\kle{0,\tau_0}$ 
\begin{align}\label{e2:rho.bound}
\rho_2(\tau)-\rho_1(\tau)\geq\tau\,\Delta\Gamma_{\mathrm{min}}\geq 0\ .
\end{align}
In appendix \ref{s:AppendixA} we show that for large $T$ 
\begin{align}\label{e2:series.bound}
\left|\sum^\infty_{n=1}\klr{L^n\chi^{(0)}_1}(T;\tau)\right|\leq C_1(T\Delta\Gamma_{\rm min})^{-1}
\left|\psi_1(T;0)\right| + C_2(T\Delta\Gamma_{\rm min})^{-2}\left|\psi_2(T;0)\right|\ ,
\end{align} 
where $C_{1,2}$ are positive constants. We get, therefore, from (\ref{e2:chi1}) with (\ref{e2:l12.bound.final}) and (\ref{e2:series.bound}) for large $T$
\begin{align}\label{2.43}
\chi_1(T;\tau)=\psi_1(T;0)\kle{1+\mcal{O}\klr{\frac{1}{T}}} + \mcal{O}\klr{\frac{1}{T}}\psi_2(T;0)\ .
\end{align}
Inserting this in (\ref{e2:Def.chi1}) we find that for large $T$ the solution of the original amplitude for the longer-lived state is
\begin{align}\label{e2:longer.lived.amp}
\begin{split}
\psi_1(T;\tau) &= \exp\kle{-\I\frac{T}{\tau_0}\varphi_1(\tau)+\I\gamma_{11}(\tau)}\\
&\times\klg{\psi_1(T;0)\kle{1+\mcal{O}\klr{\frac{1}{T}}} + \mcal{O}\klr{\frac{1}{T}}\psi_2(T;0)}\ .
\end{split}
\end{align}
For the amplitude of the shorter-lived state we obtain, in an analogous manner, from (\ref{e2:2St.ISE.2}) and (\ref{e2:longer.lived.amp}) (see appendix \ref{s:AppendixA})
\begin{align}\label{e2:shorter.lived.amp}
\begin{split}
\psi_2(T;\tau) &= \exp\kle{-\I\frac{T}{\tau_0}\varphi_1(\tau)+\I\gamma_{11}(\tau)}\klg{
\psi_1(T;0)\,\mcal{O}\klr{\frac1T}+\psi_2(T;0)\,\mcal{O}\klr{\frac1{T^2}}}\\
&+ \exp\kle{-\I\frac{T}{\tau_0}\varphi_2(\tau)+\I\gamma_{22}(\tau)}\psi_2(T;0)\ .
\end{split}
\end{align}

The interpretation of the results (\ref{e2:longer.lived.amp}) and (\ref{e2:shorter.lived.amp}) is clear. The amplitude for the longer lived state, $\alpha=1$, shows the usual dynamical and geometrical phase factors, $T\varphi_1(\tau)/\tau_0$ and $\gamma_{11}(\tau)$, respectively. Both these factors are in general complex for decaying states. The shorter lived state, $\alpha=2$, gets in general some feeding from the longer lived one. Thus the leading term of its amplitude (the first on the r.h.s. of (\ref{e2:shorter.lived.amp})) has the same exponential factor as the amplitude for $\alpha=1$. But there is a suppression by at least a factor of order $1/T$. For $\tau>0$ and large $T$ the second term on the r.h.s of (\ref{e2:shorter.lived.amp}) vanishes exponentially relative to the first term, see (\ref{e2:abs.exp}) and (\ref{e2:rho.bound}). 

In particular, if we start at $\tau=0$ with $\psi_1(T;0)\neq 0$ and $\psi_2(T;0)=0$, that is only the longer lived state is populated, the system stays in the state $\alpha=1$ up to corrections of relative order $1/T$. The amplitude evolves with the appropriate dynamical and geometrical phase factors. If, however, we start at 
$\tau=0$ with $\psi_1(T;0)=0$ and $\psi_2(T;0)\neq 0$, that is only the shorter lived state is populated, it will in general populate to a small extent the longer lived one. The latter will then feed back on the former and ``impose'' its decay properties on the system. But all this gives only amplitudes which are suppressed for large $T$. To summarise: the adiabatic theorem in the usual form can only be used for the amplitude of the longer-lived state.

\section{Several metastable and short lived states}\label{s:SeveralMetastableStates}

Here we generalise the results of section \ref{s:TwoStates} to a system having several metastable and several short lived states. In detail we suppose that the system has $N$ states where the first $M$ ones ($1\leq M<N$) are metastable. We suppose equal decay rates for the metastable states, that is
\begin{align}\label{e4:EqualDecayRates}
 \Gamma(\alpha,\tau) = \Gamma(1,\tau)
\end{align}
for $\alpha \in \klg{1,\ldots,M}$ and $\tau\in\kle{0,\tau_0}$. The remaining $N-M$ states should have substantially larger decay rates for all $\tau$: We suppose
\begin{align}\label{e4:LargerDecayRates}
\Gamma(\beta,\tau) - \Gamma(\alpha,\tau) \geq \Delta\Gamma_{\mathrm{min}} > 0
\end{align}
for all $\alpha,\beta$ with $1\leq\alpha\leq M$ and $M+1\leq\beta\leq N$ and all $\tau\in\kle{0,\tau_0}$.
Here $\Delta\Gamma_{\mathrm{min}}$ is a fixed constant.

The problem is again to solve (\ref{e2:SE.final}) for large $T$. As in section \ref{s:TwoStates} it is convenient to transform (\ref{e2:SE.final}) into an integral equation. We use the same notation as in (\ref{e2:Def.a})-(\ref{e2:Im.varphi}) but now for $1\leq\alpha,\beta\leq N$. With the initial conditions $\psi_\alpha(T;0)$ we obtain from (\ref{e2:SE.final})
\begin{align}\label{e4:psi.a}
\begin{split}
\psi_\alpha(T;\tau) &= \exp\kle{-\I\frac{T}{\tau_0}\varphi_\alpha(\tau) + \I\gamma_{\alpha\alpha}(\tau)}\\
&\times \klg{\psi_\alpha(T;0) 
+ \int_0^{\tau}\mrmd\tau'\ \exp\kle{\I\frac{T}{\tau_0}\varphi_\alpha(\tau') - \I\gamma_{\alpha\alpha}(\tau')}
\sum_{\beta\neq\alpha}\I a_{\alpha\beta}(\tau')\psi_\beta(T;\tau')}\ ,\\
&\ \quad(1 \leq \alpha,\beta \leq N)\ .
\end{split}
\end{align}

In the following we use matrix notation. We set for the metastable states
\begin{align}\label{e4:psi.metastable}
\psi_\alpha(T;\tau) &= \exp\kle{-\I\frac{T}{\tau_0}\varphi_\alpha(\tau) 
	+ \I\gamma_{\alpha\alpha}(\tau)}\chi_\alpha(T;\tau)\ ,\quad(1 \leq \alpha \leq M)\ ,\\[1mm] \label{e4:chi.vector}
\chi(T;\tau) &= 
\begin{pmatrix}
\chi_1(T;\tau)\\
\vdots\\
\chi_M(T;\tau)
\end{pmatrix}\ .
\end{align}
For the short lived states we set with $\rho_1(\tau)$ from (\ref{e2:Def.rho})
\begin{align}\label{e4:psi.unstable}
\psi_\beta(T;\tau) &= \exp\kle{-\frac{T}{2\tau_0}\rho_1(\tau)}\xi_\beta(T;\tau)\ ,\quad (M+1\leq\beta\leq N)\ ,\\[1mm]
\label{e4:xi.vector}
\xi(T;\tau) &=
\begin{pmatrix}
\xi_{M+1}(T;\tau)\\
\vdots\\
\xi_N(T;\tau)
\end{pmatrix}\ .
\end{align}
We have for $\tau=0$
\begin{align}\label{e4:chi.inital}
\chi(T;0) &= \begin{pmatrix}
\psi_1(T;0)\\
\vdots\\
\psi_M(T;0)
\end{pmatrix}
\equiv
\chi^{(0)}(T;\tau)\ ,\\[1mm] \label{e4:xi.initial}
\xi(T;0) &= 
\begin{pmatrix}
\psi_{M+1}(T;0)\\
\vdots\\
\psi_N(T;0)
\end{pmatrix}
\equiv
\xi^{(0)}(T;\tau)\ .
\end{align}
Furthermore we define
\begin{align}\label{e4:tilde.xi}
\tilde\xi^{(0)}(T;\tau) &= 
\begin{pmatrix}
\tilde\xi^{(0)}_{M+1}(T;\tau)\\
\vdots\\
\tilde\xi^{(0)}_N(T;\tau)
\end{pmatrix}\ ,\\[1mm] \label{e4:tilde.xi.comp}
\begin{split}
\tilde\xi_\alpha^{(0)}(T;\tau) &= \exp\kle{-\I\frac{T}{\tau_0}\varphi_\alpha(\tau) + \frac{T}{2\tau_0}\rho_1(\tau)
+ \I\gamma_{\alpha\alpha}(\tau)}\psi_\alpha(T;0)\ ,\\
\quad &\ \quad(M+1\leq\alpha\leq N)\ .
\end{split}
\end{align}

The integral equations (\ref{e4:psi.a}) can then be written in matrix form as follows
\begin{align}\label{e4:chi.equation}
\chi &= \chi^{(0)} + L^{(1,1)}\chi + L^{(1,2)}\xi\ ,\\ \label{e4:xi.equation}
\xi  &= \tilde\xi^{(0)} + L^{(2,1)}\chi + L^{(2,2)}\xi\ .
\end{align}
Here the $L^{(i,j)}$ are integral operators (generalising (\ref{e2:Def.l12}) and (\ref{e2:Def.L})) which are given explicitly in appendix \ref{s:AppendixB}.

The formal solution of (\ref{e4:chi.equation}) and (\ref{e4:xi.equation}) is easily written down. We get from (\ref{e4:xi.equation})
\begin{align}\label{e4:solution.step1}
\klr{1-L^{(2,2)}}\xi &= \tilde\xi^{(0)} + L^{(2,1)}\chi\ ,\\ \label{e4:solution.step2}
\xi &= \klr{1 - L^{(2,2)}}^{-1}\klr{\tilde\xi^{(0)} + L^{(2,1)}\chi}\ .
\end{align}
Inserting this in (\ref{e4:chi.equation}) gives
\begin{align}\label{e4:solution.step3}
\klg{1-L^{(1,1)} - L^{(1,2)}\klr{1-L^{(2,2)}}^{-1}L^{(2,1)}}\chi
= \chi^{(0)} + L^{(1,2)}\klr{1-L^{(2,2)}}^{-1}\tilde\xi^{(0)}\ ,
\end{align}
\begin{align}\label{e4:chi.Solution}
\chi = \klg{1-L^{(1,1)} - L^{(1,2)}\klr{1-L^{(2,2)}}^{-1}L^{(2,1)}}^{-1}
\klg{\chi^{(0)} + L^{(1,2)}\klr{1-L^{(2,2)}}^{-1}\tilde\xi^{(0)}}\ .
\end{align}
This is the exact solution. In appendix \ref{s:AppendixB} we show that for large $T$ we get from (\ref{e4:chi.Solution}) the result
\begin{align}\label{e4:chi.Solution.large.T}
\chi &= \klr{1 - L^{(1,1)}}^{-1}\chi^{(0)}
+{\cal O}\left(\frac{1}{T}\right) \Vert\chi^{(0)}\Vert+{\cal O}\left(\frac{1}{T}\right)\Vert\xi^{(0)}\Vert\ ,
\end{align}
or written in another form
\begin{align}\label{e4:chi.U}
\chi(T,\tau) &= \hat{U}(T;\tau)\chi^{(0)}
+{\cal O}\left(\frac{1}{T}\right)\Vert\chi^{(0)}\Vert
+{\cal O}\left(\frac{1}{T}\right)\Vert\xi^{(0)}\Vert\ ,
\end{align}
see (\ref{B.76}) and (\ref{B.77}). Here the norms $\Vert\chi^{(0)}\Vert$, $\Vert\xi^{(0)}\Vert$ and the $M\times M$ matrix $\hat{U}$ are defined in (\ref{eB:norm.zeta}), (\ref{eB:norm.eta}) and (\ref{eB:Def.U}), respectively. 

The solution for $\xi(T;\tau)$ is given in (\ref{e4:solution.step2}). The term $(1-L^{(2,2)})^{-1}L^{(2,1)}\chi$ can be estimated using (\ref{B.52d}) and we find
\begin{align}\label{e4:Est.Chi}
\Vert(1-L^{(2,2)})^{-1}L^{(2,1)}\chi\Vert
\leq {\cal O}\klr{\frac1T}\Vert\chi\Vert\ .
\end{align}
With (\ref{e4:chi.U}), (\ref{eB:Def.U}) and (\ref{eB:norm.op.zeta}) we get from (\ref{e4:Est.Chi}) 
\begin{align}\label{e4:Est.Chi.2}
\Vert(1-L^{(2,2)})^{-1}L^{(2,1)}\chi\Vert
\leq{\cal O}\klr{\frac1T}\Vert\chi^{(0)}\Vert+
{\cal O}\klr{\frac1{T^2}}\Vert\xi^{(0)}\Vert\ .
\end{align}
The term $(1-L^{(2,2)})^{-1}\tilde{\xi}^{(0)}$ in (\ref{e4:solution.step2}) can be estimated according to (\ref{B.66b}) 
\begin{align}\label{5.22}
\Vert(1-L^{(2,2)})^{-1}\tilde{\xi}^{(0)}\Vert
\leq\hat{C}_{22}\exp
\left[-\frac{T}{2\tau_0}\Delta\Gamma_{\rm min}\tau\right]\Vert\xi^{0)}\Vert\ . 
\end{align}
Putting everything together we find
\begin{align}\label{e4:xi.erg.large.T}
\xi(T;\tau)={\cal O}\left(\frac{1}{T}\right)\Vert\chi^{(0)}\Vert+{\cal O}\left(\frac{1}{T^2}\right)\Vert\xi^{(0)}\Vert
+\exp\left[-\frac{T}{2\tau_0}\Delta\Gamma_{\rm min}\tau\right]
{\cal O}(1)\Vert\xi^{(0)}\Vert\ .
\end{align}

Now we can go back to the original amplitudes $\psi_\alpha(T;\tau)$ of (\ref{e2:SE.final}) and (\ref{e4:psi.a}). We find from (\ref{e4:psi.metastable}), (\ref{e4:chi.inital}) and (\ref{e4:chi.U}) for the amplitudes of the metastable states
\begin{align}\label{e4:metastable.amplitudes.large.T}
\begin{split}
\psi_\alpha(T;\tau) &= \exp\left[-\I\frac{T}{\tau_0}\varphi_\alpha(\tau)\right]\\
&\quad\times\klg{\sum\limits^M_{\beta=1}U^{\rm{geom}}_{\alpha\beta}(T;\tau)\chi^{(0)}_\beta
+{\cal O}\left(\frac{1}{T}\right)\Vert\chi^{(0)}\Vert+{\cal O}\left(\frac{1}{T}\right)\Vert\xi^{(0)}\Vert}\ ,
\end{split}
\end{align}
for $\alpha\in\{1,\dots, M\}$. Here the matrix of the geometric phase factors is given by 
\begin{align}\label{e4:U.geom}
\begin{split}
U^{\rm{geom}}(T;\tau) &= \Big(U^{\rm{geom}}_{\alpha\beta}(T;\tau)\Big)\ ,\\
U^{\rm{geom}}_{\alpha\beta}(T;\tau) &= \exp[\I\gamma_{\alpha\alpha}(\tau)]\hat{U}_{\alpha\beta}(T;\tau)
\end{split}
\end{align}
with $\hat{U}(T;\tau)$ given in (\ref{eB:Def.U}).
For the amplitudes of the short lived states we find from (\ref{e4:psi.unstable}), (\ref{e4:xi.initial}) and (\ref{e4:xi.erg.large.T})
\begin{align}\label{e4:shortlived.amplitudes.large.T}
\begin{split}
\psi_\beta(T;\tau) &= \exp\left[-\frac{T}{2\tau_0}\rho_1(\tau)\right]
\left\{{\cal O}\left(\frac{1}{T}\right)
\Vert\chi^{(0)}\Vert
+{\cal O}\left(\frac{1}{T^2}\right)\Vert\xi^{(0)}\Vert\right\}\\
&\quad+\exp\left[-\frac{T}{2\tau_0} \big(\rho_1(\tau)+\Delta\Gamma_{\rm min}\tau\big)\right]
{\cal O}(1)\Vert\xi^{(0)}\Vert\ ,\\
\beta &\in \{M+1,\dots, N\}\ .
\end{split}
\end{align}

The interpretation of (\ref{e4:metastable.amplitudes.large.T}) and (\ref{e4:shortlived.amplitudes.large.T}) is analogous to the one given for the two-state system in section 4. Suppose that for $t=0$ only the metastable states  are populated, that is, we have $\chi^{(0)}\neq 0$ and $\xi^{(0)}=0$. Then the amplitudes for the metastable states evolve according to (\ref{e4:metastable.amplitudes.large.T}), obtaining a dynamical phase factor but also a --- in general non abelian --- geometric phase factor given by the matrix $U^{\rm geom}(T;\tau)$. Corrections to this are suppressed by a factor $T^{-1}$ for large $T$. Looking through the formulae of appendix \ref{s:AppendixB} we see that these suppressed terms are of order $(T\Delta\Gamma_{\rm min})^{-1}$ relative to the leading term. We shall analyse the metastable states further in section \ref{s:MetastableStates}.

On the other hand, suppose that initially only the short lived states are populated, that is, we have $\chi^{(0)}=0$ and $\xi^{(0)}\neq 0$. Then the amplitudes of the short lived states have one part showing a fast decay corresponding to the decay rates of the short lived states and terms ${\cal O}(1/T^2)$ showing the decay as for the metastable states. The metastable states get in this case only amplitudes suppressed by ${\cal O}(1/T)$.

\section{The metastable states}\label{s:MetastableStates}

In this section we study the evolution of the metastable states alone. That is, we assume $T$ to be  large enough such that all terms ${\cal O}(1/T)$ can be neglected. We have then from (\ref{e4:metastable.amplitudes.large.T})
\begin{align}\label{e6:metastable.amplitudes.large.T}
\psi_\alpha(T;\tau)= \sum\limits^M_{\beta=1}U_{\alpha\beta}(T;\tau)\chi^{(0)}_\beta\ ,\qquad
\alpha\in \{1,\dots, M\}\ ,
\end{align}
where
\begin{align}\label{e6:Def.U}
\begin{split}
U(T;\tau) &= \Big(U_{\alpha\beta}(T;\tau)\Big)\\ 
&= U^{\rm dyn}(T;\tau)U^{\rm geom}(T;\tau)\ ,
\end{split}\\ \label{e6:U.dyn}
U^{\rm dyn}(T;\tau) &= \left(\exp\left[-\I\frac{T}{\tau_0}\varphi_\alpha(\tau)\right]\delta_{\alpha\beta}\right)
\end{align}
and $U^{\rm geom}(T;\tau)$ is defined in (\ref{e4:U.geom}). This gives with $\hat U$ from (\ref{eB:Def.U})
\begin{align}
U_{\alpha\beta}(T;\tau) = \exp\left[-\I\frac{T}{\tau_0}\varphi_\alpha(\tau) + \I\gamma_{\alpha\alpha}(\tau)\right]\hat U_{\alpha\beta}(T;\tau)\ .
\end{align}
From (\ref{e2:Def.gamma}), (\ref{e2:Def.varphi}), (\ref{eB:Def.U}) and (\ref{eB:U.derivative}) we find 
\begin{align}\label{e6:U.initial}
U(T;0) &= \mathbbm{1}_M\ ,\\ \label{e6:U.derivative}
\frac{\partial}{\partial\tau}U(T;\tau) &= -\I \mathscr N(T;\tau)U(T;\tau)\ ,
\end{align}
where
\begin{align}\label{e6:Def.N}
\begin{split}
\mathscr N(T;\tau) &= \Big(\mathscr N_{\alpha\beta}(T;\tau)\Big)\ ,\\
\mathscr N_{\alpha\beta}(T;\tau) &= \frac{T}{\tau_0}E(\alpha,\tau)\delta_{\alpha\beta}-a_{\alpha\beta}(\tau)\ ,\\
\alpha,\beta &\in \{1,\ldots,M\}\ .
\end{split}
\end{align}
Thus, the amplitudes for the metastable states (\ref{e6:metastable.amplitudes.large.T}) evolve according to the effective Schr\"odinger equation
\begin{align}\label{e6:eff.SG}
\I\frac{\partial}{\partial\tau}\psi_\alpha(T;\tau)=\sum\limits^M_{\beta=1}\mathscr N_{\alpha\beta}(T;\tau)\psi_\beta(T;\tau)\ ,\qquad
\alpha\in\{1,\dots,M\}\ .
\end{align}
with effective mass matrix $\mathscr N(T;\tau)$. 

Recall that we supposed equal decay rates for the metastable states. If the energy eigenvalues for these states satisfy (\ref{e2:EW.non.deg}) we can use the results of section \ref{s:StableStates}. Then the amplitudes $\psi_\alpha(T;\tau)$ (\ref{e6:eff.SG}) decouple in the evolution for $T\to \infty$ and we get up to corrections of relative order $1/T$ (see (\ref{e2:SE.Solution.Stable}))
\begin{align}\label{e6:abelian.amplitudes}
\psi_\alpha(T;\tau) &= \exp \left[-\I\frac{T}{\tau_0}\varphi_\alpha(\tau)+\I\gamma_{\alpha\alpha}(\tau)\right]
\chi^{(0)}_\alpha\ ,\\ \label{e6:abelian.U}
U(T;\tau) &= \klr{\exp\left[-\I\frac{T}{\tau_0}\varphi_\alpha(\tau)+\I\gamma_{\alpha\alpha}(\tau)\right]\delta_{\alpha\beta}}\ ,\\
\label{e6:abelian.U.geom}
U^{\rm geom}(T;\tau)&=\Big(\exp\kle{\I\gamma_{\alpha\alpha}(\tau)}\delta_{\alpha\beta}\Big)\ .
\end{align} 
That is, in this case the matrix of geometric phase factors becomes diagonal. In the general case, however, $U^{\rm geom}(T;\tau)$ will contain non-diagonal terms. 

Finally, we note that the results of this paper remain the same if the metastable states have slightly different decay rates and we consider only times $T$ for which 
\begin{align}\label{e6:condition}
\left|\big[\Gamma(\alpha,\tau)-\Gamma(\beta,\tau)\big]T\right|\ll 1
\end{align} 
for all $\alpha,\beta\in\{1,\dots,M\}$ and all $\tau\in[0,\tau_0]$.

\section{Conclusions}\label{s:Conclusions}

In this article we have analysed the temporal behaviour of a system consisting of several metastable and several short lived states. The mass matrix governing the evolution of the system was supposed to be slowly varying with time, see (\ref{e2:WW.SE})-(\ref{e2:M.tau}). The adiabatic theorem, adapted to decaying states, was shown to hold for the metastable states, see section \ref{s:SeveralMetastableStates}. The evolution of these states (see (\ref{e6:metastable.amplitudes.large.T}), (\ref{e6:Def.U})) is governed by a dynamical phase factor matrix $U^{\rm dyn}(T,\tau)$ (\ref{e6:U.dyn}) which is diagonal and a geometric phase factor matrix $U^{\rm geom}(T;\tau)$ (\ref{e4:U.geom}) which, in general, is not diagonal. 

In the accompanying paper II we apply these results to a study of the states of hydrogen and deuterium with principal quantum number $n=2$ in slowly varying external electric and magnetic fields. The $2S$ states are metastable and we shall define and discuss parity conserving and parity violating geometric phase factors for them.

\subsection*{Acknowledgements}

The authors thank M. DeKieviet, D. Dubbers and U. Jentschura for many useful discussions. This work was supported by Deutsche Forschungsgemeinschaft under project No. NA\,296/3-1.


\appendix

\titlelabel{Appendix \thetitle}

\section{}\label{s:AppendixA}

\renewcommand{\theequation}{\thesection.\arabic{equation}}

Here we give the details of the estimates (\ref{e2:series.bound}) and (\ref{e2:shorter.lived.amp}) for the two-state system discussed in section \ref{s:TwoStates}.

We start with the proof of (\ref{e2:series.bound}). The operator $L$ is defined in (\ref{e2:Def.L}). For any continuous function $\tau\rightarrow\zeta(\tau)$ on the interval $\kle{0,\tau_0}$ we have with (\ref{e2:Decay.Assumption}) and (\ref{e2:abs.exp})
\begin{align}\label{eA:L.bound}
\begin{split}
\big|(L\zeta)(\tau)\big| &\leq
\int^\tau_0 \mrmd\tau_1~\exp\kle{-\frac{T}{2\tau_0}\big(\rho_2(\tau_1)-\rho_1(\tau_1)\big)}
\left|\exp\kle{-\I\gamma_{11}(\tau_1)+\I\gamma_{22}(\tau_1)}\right|\,|a_{12}(\tau_1)|\\
&\quad\times \int^{\tau_1}_0 \mrmd\tau_2~\exp\kle{\frac{T}{2\tau_0}\big(\rho_2(\tau_2)-\rho_1(\tau_2)\big)}
\big[\Gamma(2,\tau_2)-\Gamma(1,\tau_2)\big]\\
&\quad\quad\times \Delta\Gamma_{\mathrm{min}}^{-1}
\left|\exp\kle{\I\gamma_{11}(\tau_2)-\I\gamma_{22}(\tau_2)}\right|\,|a_{21}(\tau_2)|
\max_{\tau'\in\kle{0,\tau_0}}|\zeta(\tau')|\ .
\end{split}
\end{align}
Analogously to (\ref{e2:Gmin.bound}) we define
\begin{align}\label{eA:Gmin.bound}
c_{21} = \max_{\tau_2\in \kle{0,\tau_0}}\klg{\tau_0\left|\exp\kle{\I\gamma_{11}(\tau_2)-\I\gamma_{22}(\tau_2)}\right|\,
|a_{21}(\tau_2)|}
\end{align}
where $c_{21}$ is a finite, non-negative constant. We get then with (\ref{e2:Gmin.bound}) and (\ref{eA:Gmin.bound})
\begin{align}\label{A.3}
\begin{split}
\big|(L\zeta)(\tau)\big| &\leq
c_{12}c_{21}\,\tau^{-2}_0\Delta\Gamma_{\mathrm{min}}^{-1}\,\max_{\tau'\in\kle{0,\tau_0}}\left|\zeta(\tau')\right|
\int^\tau_0 \mrmd\tau_1~\exp\kle{-\frac{T}{2\tau_0}\big(\rho_2(\tau_1)-\rho_1(\tau_1)\big)}\\
&\quad\times \int^{\tau_1}_0 \mrmd\tau_2~\exp\kle{\frac{T}{2\tau_0}\big(\rho_2(\tau_2)-\rho_1(\tau_2)\big)}
\frac{\mrmd}{\mrmd\tau_2}\big(\rho_2(\tau_2)-\rho_1(\tau_2)\big)\\
&= c_{12}c_{21}\, \tau^{-2}_0\Delta\Gamma_{\mathrm{min}}^{-1}\,\max_{\tau'\in\kle{0,\tau_0}}\left|\zeta(\tau')\right|\\
&\quad\times\frac{2\tau_0}{T}
\int^\tau_0\mrmd\tau_1~\klg{1-\exp\kle{-\frac{T}{2\tau_0}\big(\rho_2(\tau_1)-\rho_1(\tau_1)\big)}}\\
&\leq c_{12}c_{21}\,\Delta\Gamma_{\mathrm{min}}^{-1}\,\frac{2}{T}
\max_{\tau'\in\kle{0,\tau_0}}\left|\zeta(\tau')\right|\ .
\end{split}
\end{align}
Setting 
\begin{align}\label{A.4}
\tilde{\tau}_0 = 2c_{12}c_{21}\,\Delta\Gamma_{\mathrm{min}}^{-1}
\end{align}
we have thus
\begin{align}\label{A.5}
\big|(L\zeta)(\tau)\big|\leq\frac{\tilde{\tau}_0}{T}\max_{\tau'\in\kle{0,\tau_0}}\left|\zeta(\tau')\right|
\end{align}
and by straightforward iteration
\begin{align}\label{A.6}
\big|(L^n\zeta)(\tau)\big|\leq\klr{\frac{\tilde{\tau}_0}{T}}^n
\max_{\tau'\in\kle{0,\tau_0}}\left|\zeta(\tau')\right|\ .
\end{align}
For large enough $T$ we have certainly $T>2\tilde{\tau}_0$ and we get
\begin{align}\label{A.7}
\begin{split}
\left|\sum^\infty_{n=1}(L^n\zeta)(\tau)\right| &\leq
\sum^\infty_{n=1}\klr{\frac{\tilde{\tau}_0}{T}}^n\max_{\tau'\in\kle{0,\tau_0}}\left|\zeta(\tau')\right|\\
&= \frac{\tilde{\tau}_0}{T}\klr{1-\frac{\tilde{\tau}_0}{T}}^{-1}
\max_{\tau'\in\kle{0,\tau_0}}\left|\zeta(\tau')\right|\\
&\leq \frac{2\tilde{\tau}_0}{T}\max_{\tau'\in\kle{0,\tau_0}}\left|\zeta(\tau')\right|\ .
\end{split}
\end{align}
Replacing here $\zeta(\tau)$ by $\chi_1^{(0)}(T;\tau)$ we get with (\ref{e2:Def.chi1.0}) and (\ref{e2:l12.bound.final})
\begin{align}\label{A.8}
\begin{split}
\left|\sum^\infty_{n=1}(L^n\chi^{(0)}_1)(T;\tau)\right|
&\leq \frac{2\tilde{\tau}_0}{T}\max_{\tau'\in\kle{0,\tau_0}}\left|\chi^{(0)}_1(T,\tau')\right|\\
&\leq \frac{2\tilde{\tau}_0}{T}\klg{\left|\psi_1(T;0)\right|+\mcal{O}\klr{\frac{1}{T}}\left|\psi_2(T;0)\right|}
\end{split}
\end{align}
which proves (\ref{e2:series.bound}).

For the proof of (\ref{e2:shorter.lived.amp}) we start from (\ref{e2:2St.ISE.2}) and (\ref{e2:Def.chi1}) to get
\begin{align}\label{A.9}
\psi_2(T;\tau)=\psi^{(1)}_2(T;\tau)+\psi^{(2)}_2(T;\tau)\ ,
\end{align}
where
\begin{align}\label{A.10}
\psi^{(1)}_2(T;\tau) &=
\exp\kle{-\I\frac{T}{\tau_0}\varphi_1(\tau)+\I\gamma_{11}(\tau)}\chi^{(1)}_2(T;\tau)\ ,\\ \label{A.11}
\begin{split}
\chi^{(1)}_2(T;\tau) &=
\exp\kle{-\I\frac{T}{\tau_0}\big(\varphi_2(\tau)-\varphi_1(\tau)\big)}
\exp\kle{-\I\gamma_{11}(\tau)+i\gamma_{22}(\tau)}\\
&\quad\times \int^\tau_0\mrmd\tau_1~\exp\kle{\I\frac{T}{\tau_0}\big(\varphi_2(\tau_1)-\varphi_1(\tau_1)\big)}
\exp\kle{\I\gamma_{11}(\tau_1)-\I\gamma_{22}(\tau_1)} \I a_{21}(\tau_1)\\
&\quad\times \chi_1(T;\tau_1)\ ,
\end{split}\\ \label{A.12}
\psi^{(2)}_2(T;\tau) &= \exp\kle{-\I\frac{T}{\tau_0}\varphi_2(\tau)+\I\gamma_{22}(\tau)}\psi_2(T;0)\ .
\end{align}
Using the same techniques as for $(L\zeta)(\tau)$ we find with $c_{21}$ from (\ref{eA:Gmin.bound}) and with (\ref{2.43})
\begin{align}\label{A.13}
\begin{split}
\left|\chi^{(1)}_2(T;\tau)\right| &\leq 
c_{21}(\tau_0\Delta\Gamma_{\rm min})^{-1}\max_{\tau'\in\kle{0,\tau_0}}
\left|\exp\kle{-\I\gamma_{11}(\tau')+\I\gamma_{22}(\tau')}\right|\\ 
&\quad\times\exp\kle{-\frac{T}{2\tau_0}\big(\rho_2(\tau)-\rho_1(\tau)\big)}\\
&\quad\times
\int^\tau_0\mrmd\tau_1~\exp\kle{\frac{T}{2\tau_0}\big(\rho_2(\tau_1)-\rho_1(\tau_1)\big)}
\frac{\mrmd}{\mrmd\tau_1}\big(\rho_2(\tau_1)-\rho_1(\tau_1)\big)\\
&\quad\times
\klg{\left|\psi_1(T;0)\right|\kle{1+\mcal{O}\klr{\frac1T}} + \mcal{O}\klr{\frac1T}\left|\psi_2(T;0)\right|}\\
&\leq \frac{2}{T}\Delta\Gamma_{\rm min}^{-1}
c_{21}\max_{\tau'\in\kle{0,\tau_0}}
\left|\exp\kle{-\I\gamma_{11}(\tau')+\I\gamma_{22}(\tau')}\right|\\
&\quad\times \klg{\left|\psi_1(T;0)\right|\kle{1+\mcal{O}\klr{\frac1T}} 
+ \mcal{O}\klr{\frac1T}\left|\psi_2(T;0)\right|}\ .
\end{split}
\end{align}
Inserting (\ref{A.10})-(\ref{A.13}) in (\ref{A.9}) proves (\ref{e2:shorter.lived.amp}).

To summarise, we find the following results where for later use in II we write out the nominal order of magnitude of the first correction terms
\begin{align}\label{eA:estPsiLong}
\begin{split}
\psi_1(T;\tau) &= \exp\kle{-\I\frac T{\tau_0}\varphi_1(\tau) + \I\gamma_{11}(\tau)}\\
&\times \klg{\psi_1(T;0)\kle{1+\frac{4c_{12}c_{21}}{T\,\Delta\Gamma_{\mathrm{min}}}\,\mcal O(1)}
+ \frac{2c_{12}}{T\,\Delta\Gamma_{\mathrm{min}}}\,\mcal O(1)\psi_2(T;0)}\ ,
\end{split}\\ \label{eA:estPsiShort}
\begin{split}
\psi_2(T;\tau) &= \exp\kle{-\I\frac{T}{\tau_0}\varphi_1(\tau) + \I\gamma_{11}(\tau)}\\
&\times \klg{\frac{2c_{21}}{T\,\Delta\Gamma_{\mathrm{min}}}\,\mcal O(1)\psi_1(T;0)
+ \frac{4c_{21}c_{12}}{(T\,\Delta\Gamma_{\mathrm{min}})^2}\,\mcal O(1)\psi_2(T;0)}\\
&+ \exp\kle{-\I\frac T{\tau_0}\varphi_2(\tau) + \I\gamma_{22}(\tau)}\psi_2(T;0)\ .
\end{split}
\end{align}

\section{}\label{s:AppendixB}

Here we give the details of the calculations in section \ref{s:SeveralMetastableStates}.

Inserting (\ref{e4:psi.metastable})-(\ref{e4:xi.vector}) in (\ref{e4:psi.a}) we get (\ref{e4:chi.equation})
and (\ref{e4:xi.equation}) where we define the operators $L^{(i,j)}$ ($1\leq i,j\leq 2$) as follows. Let
\begin{align}\label{eB:Def.zeta}
\tau\to\zeta(\tau) =
\begin{pmatrix}
\zeta_1(\tau)\\
\vdots\\
\zeta_M(\tau)
\end{pmatrix}\ ,
\end{align}
and
\begin{align}\label{eB:Def.eta}
\tau\to\eta(\tau) =
\begin{pmatrix}
\eta_{M+1}(\tau)\\
\vdots\\
\eta_N(\tau)
\end{pmatrix}
\end{align}
be continuous vector functions for $\tau\in\kle{0,\tau_0}$. We define the norm for such functions as
\begin{align}\label{eB:norm.zeta}
\Vert\zeta\Vert &= \max_{\substack{1\leq\alpha\leq M,\\ \tau\in\kle{0,\tau_0}}}|\zeta_\alpha(\tau)|\ ,\\
\label{eB:norm.eta}
\Vert\eta\Vert &= \max_{\substack{M+1\leq\beta\leq N,\\ \tau\in\kle{0,\tau_0}}}|\eta_\beta(\tau)|\ .
\end{align}
The operators $L^{(i,j)}$ are defined as
\begin{flalign}\label{eB:L11}
\qquad\begin{split}
\klr{L^{(1,1)}\zeta}_\alpha(\tau) &= \I\int_0^\tau\mrmd\tau'\ \exp\kle{\I\frac{T}{\tau_0}\varphi_\alpha(\tau')
-\I\gamma_{\alpha\alpha}(\tau')}\\
&\quad\times \sum_{\beta=1}^M a_{\alpha\beta}(\tau')\klr{1-\delta_{\alpha\beta}}
\exp\kle{-\I\frac{T}{\tau_0}\varphi_\beta(\tau') + \I\gamma_{\beta\beta}(\tau')}\zeta_\beta(\tau')\ ,
\end{split}&&\\ \nonumber
(1 &\leq \alpha\leq M)\ ,&&
\end{flalign}\begin{flalign}
\label{eB:L12}
\qquad\begin{split}
\klr{L^{(1,2)}\eta}_\alpha(\tau) &= \I\int_0^\tau\mrmd\tau'\ \exp\kle{\I\frac{T}{\tau_0}\varphi_\alpha(\tau')
-\I\gamma_{\alpha\alpha}(\tau')}\\
&\quad\times \sum_{\beta=M+1}^N a_{\alpha\beta}(\tau')\exp\kle{-\frac{T}{2\tau_0}\rho_1(\tau')}\eta_\beta(\tau')\ ,
\end{split}&&\\ \nonumber
(1 &\leq \alpha\leq M)\ ,&&
\end{flalign}\begin{flalign}
\label{eB:L21}
\qquad\begin{split}
\klr{L^{(2,1)}\zeta}_\alpha(\tau) &= \exp\kle{-\I\frac{T}{\tau_0}\varphi_\alpha(\tau) + \frac{T}{2\tau_0}\rho_1(\tau)
+ \I\gamma_{\alpha\alpha}(\tau)}\\
&\quad\times \I\int_0^\tau\mrmd\tau'\ \exp\kle{\I\frac{T}{\tau_0}\varphi_\alpha(\tau')-\I\gamma_{\alpha\alpha}(\tau')}\\
&\quad\times \sum_{\beta=1}^M a_{\alpha\beta}(\tau')\exp\kle{-\I\frac{T}{\tau_0}\varphi_\beta(\tau')
+\I\gamma_{\beta\beta}(\tau')}\zeta_\beta(\tau')\ ,
\end{split}&&\\ \nonumber
(M+1 &\leq \alpha\leq N)\ ,&&
\end{flalign}\begin{flalign}
\label{eB:L22}
\qquad\begin{split}
\klr{L^{(2,2)}\eta}_\alpha(\tau) &= \exp\kle{-\I\frac{T}{\tau_0}\varphi_\alpha(\tau) + \frac{T}{2\tau_0}\rho_1(\tau)
+ \I\gamma_{\alpha\alpha}(\tau)}\\
&\quad\times \I\int_0^\tau\mrmd\tau'\ \exp\kle{\I\frac{T}{\tau_0}\varphi_\alpha(\tau')-\I\gamma_{\alpha\alpha}(\tau')}\\
&\quad\times \sum_{\beta=M+1}^N a_{\alpha\beta}(\tau')\klr{1-\delta_{\alpha\beta}}\exp\kle{-\frac{T}{2\tau_0}\rho_1(\tau')}\eta_\beta(\tau')\ .
\end{split}&&\\ \nonumber
(M+1 &\leq \alpha\leq N)\ .&&
\end{flalign}

Our aim is to derive estimates similar to (\ref{e2:l12.bound.final}) and (\ref{A.3})-(\ref{A.8}) in order to prove (\ref{e4:chi.Solution.large.T}).
Let us first recall that we are supposing all functions of $\tau$ to be continuous for $\tau\in\kle{0,\tau_0}$. Thus we can define the following dimensionless nonnegative and finite constants
\begin{align}\label{eB:Def.C11}
\begin{split}
C_{11} &= \ \ \max_{\substack{1\leq\alpha\leq M,\\ 1\leq\beta\leq M,\\ \tau\in\kle{0,\tau_0}}}\ \ 
\tau_0\Big|a_{\alpha\beta}(\tau)\klr{1-\delta_{\alpha\beta}}
\exp\kle{-\I\gamma_{\alpha\alpha}(\tau) + \I\gamma_{\beta\beta}(\tau)}\Big|\ ,
\end{split}\\ \label{eB:Def.C12}
\begin{split}
C_{12} &= \max_{\substack{1\leq\alpha\leq M,\\ M+1\leq\beta\leq N,\\ \tau\in\kle{0,\tau_0}}} 
\tau_0\Big|a_{\alpha\beta}(\tau)
\exp\kle{-\I\gamma_{\alpha\alpha}(\tau) + \I\gamma_{\beta\beta}(\tau)}\Big|\ ,
\end{split}
\end{align}
\begin{align}\label{eB:Def.C21}
\begin{split}
C_{21} &= \max_{\substack{M+1\leq\alpha\leq N,\\ 1\leq\beta\leq M,\\ \tau\in\kle{0,\tau_0}}} 
\tau_0\Big|a_{\alpha\beta}(\tau)
\exp\kle{-\I\gamma_{\alpha\alpha}(\tau) + \I\gamma_{\beta\beta}(\tau)}\Big|\ ,
\end{split}\\ \label{eB:Def.C22}
\begin{split}
C_{22} &= \max_{\substack{M+1\leq\alpha\leq N,\\ M+1\leq\beta\leq N,\\ \tau\in\kle{0,\tau_0}}} 
\tau_0\Big|a_{\alpha\beta}(\tau)\klr{1-\delta_{\alpha\beta}}
\exp\kle{-\I\gamma_{\alpha\alpha}(\tau) + \I\gamma_{\beta\beta}(\tau)}\Big|\ .
\end{split}
\end{align}

Now we consider the solution for $\chi$ given in (\ref{e4:chi.Solution}). With simple algebra we can rewrite it as follows
\begin{align}\label{eB:chi.Solution}
\begin{split}
\chi &= \klr{1-L^{(1,1)}}^{-1}\klr{1-\tilde L}^{-1}\klg{\chi^{(0)} + L^{(1,2)}\klr{1-L^{(2,2)}}^{-1}\tilde\xi^{(0)}}\\
&= \klr{1-L^{(1,1)}}^{-1}\klr{\sum_{k=0}^\infty \tilde L^k}\klg{\chi^{(0)} 
+ L^{(1,2)}\klr{1-L^{(2,2)}}^{-1}\tilde\xi^{(0)}}\ ,
\end{split}
\end{align}
where
\begin{align}\label{eB:Def.L}
\tilde L = L^{(1,2)}\klr{1-L^{(2,2)}}^{-1}L^{(2,1)}\klr{1-L^{(1,1)}}^{-1}\ .
\end{align}

We study first the operator $\klr{1-L^{(1,1)}}^{-1}$. Let
\begin{align}\label{eB:Def.zetac}
\zeta_c =
\begin{pmatrix}
\zeta_1\\ \vdots\\ \zeta_M
\end{pmatrix}
\end{align}
be a constant vector. We define a $M\times M$ matrix function $\hat U(T;\tau)$ by
\begin{align}\label{eB:Def.U}
\begin{split}
\hat U(T;\tau)\zeta_c &= \klr{\klr{1-L^{(1,1)}}^{-1}\zeta_c}(\tau)\\
&= \sum_{r=0}^\infty\klr{\klr{L^{(1,1)}}^r\zeta_c}(\tau)\\
&= \zeta_c + \klr{L^{(1,1)}\sum_{r=0}^{\infty}\klr{L^{(1,1)}}^r\zeta_c}(\tau)\ .
\end{split}
\end{align}
From (\ref{eB:L11}) and (\ref{eB:Def.U}) we find
\begin{align}\label{eB:U.derivative}
\begin{split}
\frac{\partial}{\partial\tau}\hat U(T;\tau)\zeta_c &= 
\frac{\partial}{\partial\tau}\klg{\zeta_c + \klr{L^{(1,1)}\sum_{r=0}^{\infty}\klr{L^{(1,1)}}^r\zeta_c}(\tau)}\\
&= A(T;\tau)\hat U(T;\tau)\zeta_c
\end{split}
\end{align}
where
\begin{align}\label{eB:Def.A}
\begin{split}
A(T;\tau) &= \Big(A_{\alpha\beta}(T;\tau)\Big)\ ,\\
A_{\alpha\beta}(T;\tau) &= \exp\kle{\I\frac{T}{\tau_0}\varphi_\alpha(\tau) - \I\gamma_{\alpha\alpha}(\tau)}
\I a_{\alpha\beta}(\tau)\klr{1-\delta_{\alpha\beta}}\\
&\times \exp\kle{-\I\frac{T}{\tau_0}\varphi_\beta(\tau) + \I\gamma_{\beta\beta}(\tau)}\ ,\\
(1\leq &~\alpha,\beta\leq M)\ .
\end{split}
\end{align}
We have
\begin{align}\label{eB:U.props}
\begin{split}
\hat U(T;0) &= \mathbbm1_M\ ,\\
\det \hat U(T;0) &= 1\ ,
\end{split}\\ \label{eB:Tr.A}
\Tr A(T;\tau) &= 0\ ,\ \text{for all $\tau\in\kle{0,\tau_0}$.}
\end{align}
We also have the relation
\begin{align}\label{eB:det.U.derivative}
\frac{\partial}{\partial\tau}\det\hat U(T;\tau) = \Tr A(T;\tau)\det\hat U(T;\tau)
\end{align}
following from (\ref{eB:U.derivative}). We find, therefore, with (\ref{eB:Tr.A}) and (\ref{eB:U.props}) 
\begin{align}\label{eB:det.U}
\det\hat U(T;\tau) = \det\hat U(T;0) = 1
\end{align}
for all $\tau\in\kle{0,\tau_0}$. Thus $\hat{U}(T;\tau)$ is never singular and has an inverse $\hat{U}^{-1}(T;\tau)$ for all $\tau\in\kle{0,\tau_0}$. For the inverse  we have
\begin{align}\label{eB:U.inv.derivative}
\frac{\partial}{\partial\tau}\hat{U}^{-1}(T;\tau) = - \hat{U}^{-1}(T;\tau)A(T;\tau)\ .
\end{align}

The matrix elements $A_{\alpha\beta}(T;\tau)$ (\ref{eB:Def.A}) are bounded as we see from (\ref{eB:Def.C11}),
(\ref{e2:abs.exp}) and (\ref{e4:EqualDecayRates}):
\begin{align}\label{eB:A.bound}
\left|A_{\alpha\beta}(T;\tau)\right| \leq \frac{C_{11}}{\tau_0}
\end{align}
for all $\tau\in\kle{0,\tau_0}$. We show now that also the matrix elements $\hat U_{\alpha\beta}(T;\tau)$ and $\hat U^{-1}_{\alpha\beta}(T;\tau)$ are bounded. Indeed, consider
\begin{align}\label{eB:Def.B}
B(T;\tau) = \hat U(T;\tau)\hat U^\dagger(T;\tau)\ .
\end{align}
We can diagonalise this matrix
\begin{align}\label{eB:B.diag}
S(T;\tau) B(T;\tau) S^\dagger(T;\tau) &= \diag\klr{b_1(T;\tau),\ldots,b_M(T;\tau)} \equiv B_{\mathrm{diag}}(T;\tau)\ ,\\
S(T;\tau)S^\dagger(T;\tau) &= \mathbbm1\ .
\end{align}
The eigenvalues $b_\alpha(T;\tau)$ must be positive
\begin{align}\label{eB:b.pos}
b_\alpha(T;\tau) > 0\ ,\ \text{for all $\tau\in\kle{0,\tau_0},$}\ (\alpha=1,\ldots,M)\ .
\end{align}
Now we consider $\Tr B(T;\tau)$ and find from (\ref{eB:U.derivative})
\begin{align}\label{eB:Tr.B.derivative}
\begin{split}
\frac{\partial}{\partial\tau}\Tr B(T;\tau) &= \Tr\klg{A(T;\tau)B(T;\tau) + B(T;\tau)A^\dagger(T;\tau)}\\
&= \Tr\klg{B_{\mathrm{diag}}(T;\tau)\kle{S(T;\tau)\klr{A(T;\tau)+A^\dagger(T;\tau)}S^\dagger(T;\tau)}}\\
&= \sum_{\alpha=1}^M b_\alpha(T;\tau) A'_{\alpha\alpha}(T;\tau)\ ,
\end{split}
\end{align}
where
\begin{align}\label{eB:A.prime}
A'_{\alpha\alpha}(T;\tau) = \kle{S(T;\tau)\klr{A(T;\tau)+A^\dagger(T;\tau)}S^\dagger(T;\tau)}_{\alpha\alpha}\ .
\end{align}
Suppressing the arguments $(T;\tau)$ we have
\begin{align}\label{eB:A.prime.diagelem}
A'_{\alpha\alpha} &= \sum_{\beta,\gamma=1}^M\klr{A_{\beta\gamma} + A_{\gamma\beta}^*}S_{\alpha\beta}S_{\alpha\gamma}^*\ ,\\ \label{eB:A.prime.squared}
\begin{split}
\left|A'_{\alpha\alpha}\right|^2 &= 
\left|\sum_{\beta,\gamma=1}^M\klr{A_{\beta\gamma} + A_{\gamma\beta}^*}S_{\alpha\beta}S_{\alpha\gamma}^*\right|^2\\
&\leq \klr{\sum_{\beta,\gamma=1}^M\left|A_{\beta\gamma} + A_{\gamma\beta}^*\right|^2}
\sum_{\beta,\gamma=1}^M\klr{S_{\alpha\beta}S^*_{\alpha\gamma}S^*_{\alpha\beta}S_{\alpha\gamma}}\\
&= \sum_{\beta,\gamma=1}^M\left|A_{\beta\gamma} + A^*_{\gamma\beta}\right|^2\\
&\leq 4M^2\klr{\frac{C_{11}}{\tau_0}}^2\ .
\end{split}
\end{align}
Thus we find
\begin{align}\label{eB:abs.A.prime}
\left|A'_{\alpha\alpha}\right| \leq 2M\frac{C_{11}}{\tau_0}
\end{align}
and from (\ref{eB:Tr.B.derivative})
\begin{align}\label{eB:Tr.B.der.bound}
\begin{split}
\frac{\partial}{\partial\tau}\Tr B(T;\tau) &\leq 2M\frac{C_{11}}{\tau_0}\sum_{\alpha=1}^M b_\alpha(T;\tau)\\
&= 2M\frac{C_{11}}{\tau_0}\Tr B(T;\tau)\ .
\end{split}
\end{align}
From (\ref{eB:Tr.B.der.bound}) and the initial condition (\ref{eB:U.props}) we see that
\begin{align}\label{eB:Tr.U.bound}
\Tr\klg{\hat U(T;\tau)\hat U^\dagger(T;\tau)} \leq M\exp\klr{2MC_{11}}
\end{align}
for all $\tau\in\kle{0,\tau_0}$. This shows that all matrix elements of $\hat U(T;\tau)$ stay bounded and, in particular, have no terms growing exponentially with $T$. Using (\ref{eB:U.inv.derivative}) it is easy to show that the same holds for all matrix elements of $\hat U^{-1}(T;\tau)$.

Consider now an arbitrary vector function $\tau\to\zeta(\tau)$ as in (\ref{eB:Def.zeta}) and set
\begin{align}\label{eB:tilde.zeta}
\begin{split}
\tilde\zeta_T(\tau) &= \klr{\klr{1-L^{(1,1)}}^{-1}\zeta}(\tau)\\ 
&= \zeta(\tau) + \klr{L^{(1,1)}\tilde\zeta_T}(\tau)\ .
\end{split}
\end{align}
We have then
\begin{align}\label{eB:tilde.zeta.initial}
\tilde\zeta_T(0) &= \zeta(0)\ ,\\ \label{eB:tilde.zeta.derivative}
\frac{\partial}{\partial\tau}\tilde\zeta_T(\tau) &= \frac{\partial\zeta(\tau)}{\partial\tau}
+ A(T;\tau)\tilde\zeta_T(\tau)\ .
\end{align}
The solution is
\begin{align}\label{eB:tilde.zeta.solution}
\tilde\zeta_T(\tau) = \zeta(\tau) + \hat U(T;\tau)\int_0^\tau\mrmd\tau'\ \hat U^{-1}(T;\tau')
A(T;\tau')\zeta(\tau')\ .
\end{align}
Since the matrix elements of $A(T;\tau)$, $\hat U(T;\tau)$ and $\hat U^{-1}(T;\tau)$ are all bounded, see (\ref{eB:A.bound})ff.,
we see that with the norm as in (\ref{eB:norm.zeta}) we get
\begin{align}\label{eB:norm.tilde.zeta}
\Vert\tilde\zeta_T\Vert \leq \tilde C\Vert\zeta\Vert\ .
\end{align}
Here $\tilde C$ is a constant. That is, we have from (\ref{eB:tilde.zeta})
\begin{align}\label{eB:norm.op.zeta}
\Vert\klr{1-L^{(1,1)}}^{-1}\zeta\Vert \leq \tilde C\Vert\zeta\Vert\ .
\end{align}

The next term to consider is the following operator occurring in (\ref{eB:Def.L})
\begin{align}\label{eB:K.series}
\begin{split}
L^{(1,2)}\klr{1-L^{(2,2)}}^{-1}L^{(2,1)} &= \sum_{r=0}^\infty L^{(1,2)}\klr{L^{(2,2)}}^r L^{(2,1)}\\
&= \sum_{r=0}^\infty K^{(r)}\ ,
\end{split}
\end{align}
where
\begin{align}\label{eB:Def.K}
K^{(r)} = L^{(1,2)}\klr{L^{(2,2)}}^{r}L^{(2,1)}\ .
\end{align}
Let $\zeta(\tau)$ be as in (\ref{eB:Def.zeta}). We get then
\begin{align}\label{eB:K0}
\begin{split}
\klr{K^{(0)}\zeta}_\alpha(\tau) &= \klr{L^{(1,2)}L^{(2,1)}\zeta}_\alpha(\tau)\\
&= \sum_{\beta=M+1}^N\sum_{\varkappa=1}^M\ \I\int_0^\tau\mrmd\tau_1\ \exp\kle{
\I\frac{T}{\tau_0}\varphi_\alpha(\tau_1) - \I\frac{T}{\tau_0}\varphi_\beta(\tau_1)}\\
&\quad\times 
a_{\alpha\beta}(\tau_1)\exp\kle{-\I\gamma_{\alpha\alpha}(\tau_1) + \I\gamma_{\beta\beta}(\tau_1)}\\
&\quad\times
\I\int_0^{\tau_1}\mrmd\tau_2\ \exp\kle{\I\frac{T}{\tau_0}\varphi_\beta(\tau_2)
-\I\frac{T}{\tau_0}\varphi_\varkappa(\tau_2)}\\
&\quad\times
a_{\beta\varkappa}(\tau_2)\exp\kle{-\I\gamma_{\beta\beta}(\tau_2)+\I\gamma_{\varkappa\varkappa}(\tau_2)}
\zeta_\varkappa(\tau_2)\ ,\\
(\alpha &= 1,\ldots,M)\ .
\end{split}
\end{align}
Using (\ref{e2:abs.exp}), (\ref{eB:Def.C12}) and (\ref{eB:Def.C21}) we find the following estimates
\begin{align}\label{eB:abs.K0zeta}
\begin{split}
\left|\klr{K^{(0)}\zeta}_\alpha(\tau)\right| 
&\leq \frac1{\tau_0^2}C_{12}C_{21}\Vert\zeta\Vert\Delta\Gamma^{-1}_{\mathrm{min}}
\sum_{\beta=M+1}^N\sum_{\varkappa=1}^M\int_0^\tau\mrmd\tau_1\ \exp\kle{\frac{T}{2\tau_0}
\big(\rho_1(\tau_1) - \rho_\beta(\tau_1)\big)}\\
&\quad\times
\int_0^{\tau_1}\mrmd\tau_2\ \exp\kle{\frac{T}{2\tau_0}\big(\rho_\beta(\tau_2) - \rho_1(\tau_2)\big)}
\kle{\Gamma(\beta,\tau_2) - \Gamma(1,\tau_2)}\\
&\leq M(N-M)C_{12}C_{21}\Delta\Gamma^{-1}_{\mathrm{min}}\frac2T\Vert\zeta\Vert\ ,
\end{split}\\[2mm] \label{eB:norm.K0zeta}
\Vert K^{(0)}\zeta\Vert &\leq M(N-M)C_{12}C_{21}\Delta\Gamma^{-1}_{\mathrm{min}}\frac2T\Vert\zeta\Vert\ .
\end{align}
In a similar way we get for $1\leq\alpha\leq M$
\begin{align}\label{eB:K1}
\begin{split}
\klr{K^{(1)}\zeta}_\alpha(\tau) &= \klr{L^{(1,2)}L^{(2,2)}L^{(2,1)}\zeta}_\alpha(\tau)\\
&= \sum_{\beta = M+1}^N \sum_{\gamma = M+1}^N \sum_{\varkappa = 1}^M \I \int_0^{\tau}\mrmd\tau_1\ 
\exp\kle{\I\frac{T}{\tau_0}\varphi_\alpha(\tau_1) - \I\frac{T}{\tau_0}\varphi_\beta(\tau_1)}\\
&\quad\times 
a_{\alpha\beta}(\tau_1)\exp\kle{-\I\gamma_{\alpha\alpha}(\tau_1) + \I\gamma_{\beta\beta}(\tau_1)}\\
&\quad\times\I \int_0^{\tau_1}\mrmd\tau_2\ 
\exp\kle{\I\frac{T}{\tau_0}\varphi_\beta(\tau_2) - \I\frac{T}{\tau_0}\varphi_\gamma(\tau_2)}\\
&\quad\times 
a_{\beta\gamma}(\tau_2)\klr{1-\delta_{\beta\gamma}}\exp\kle{-\I\gamma_{\beta\beta}(\tau_2) + \I\gamma_{\gamma\gamma}(\tau_2)}\\
&\quad\times \I \int_0^{\tau_2}\mrmd\tau_3\ 
\exp\kle{\I\frac{T}{\tau_0}\varphi_\gamma(\tau_3) - \I\frac{T}{\tau_0}\varphi_\varkappa(\tau_3)}\\
&\quad\times 
a_{\gamma\varkappa}(\tau_3)\exp\kle{-\I\gamma_{\gamma\gamma}(\tau_3) +  \I\gamma_{\varkappa\varkappa}(\tau_3)}\zeta_\varkappa(\tau_3)\ ,
\end{split}
\end{align}
\begin{align}\label{eB:abs.K1zeta}
\begin{split}
\left|\klr{K^{(1)}\zeta}_\alpha(\tau)\right| 
&\leq \sum_{\beta=M+1}^N \sum_{\gamma=M+1}^N \sum_{\varkappa=1}^M
\int_0^\tau\mrmd\tau_1\ \exp\kle{\frac{T}{2\tau_0}\big(\rho_1(\tau_1) - \rho_\beta(\tau_1)\big)}\frac{C_{12}}{\tau_0}\\
&\quad\times
\int_0^{\tau_1}\mrmd\tau_2\ \exp\kle{\frac{T}{2\tau_0}\big(\rho_\beta(\tau_2) - \rho_\gamma(\tau_2)\big)}
\frac{\Gamma(\beta,\tau_2) - \Gamma(1,\tau_2)}{\tau_0\Delta\Gamma_{\mathrm{min}}}C_{22}\\
&\quad\times
\int_0^{\tau_2}\mrmd\tau_3\ \exp\kle{\frac{T}{2\tau_0}\big(\rho_\gamma(\tau_3) - \rho_1(\tau_3)\big)}
\frac{\Gamma(\gamma,\tau_3) - \Gamma(1,\tau_3)}{\tau_0\Delta\Gamma_{\mathrm{min}}}C_{21}\Vert\zeta\Vert\\
&\leq M(N-M)C_{12}C_{21}\Delta\Gamma^{-1}_{\mathrm{min}}\frac2T
(N-M)C_{22}\Delta\Gamma^{-1}_{\mathrm{min}}\frac2T\Vert\zeta\Vert\ .
\end{split}
\end{align}
Thus we get
\begin{align}\label{eB:norm.K1zeta}
\Vert K^{(1)}\zeta\Vert \leq M(N-M)C_{12}C_{21}\Delta\Gamma^{-1}_{\mathrm{min}}\frac2T
(N-M)C_{22}\Delta\Gamma^{-1}_{\mathrm{min}}\frac2T\Vert\zeta\Vert\ .
\end{align}
It is easy to see that this can be generalised to
\begin{align}\label{B.51}
\begin{split}
\Vert K^{(r)}\zeta\Vert &\leq M(N-M)C_{12}C_{21}\Delta\Gamma^{-1}_{\mathrm{min}}\frac2T
\kle{(N-M)C_{22}\Delta\Gamma^{-1}_{\mathrm{min}}\frac2T}^r\Vert\zeta\Vert\ ,\\
(r &= 0,1,2,\ldots)\ .
\end{split}
\end{align}

Now we can go back to (\ref{eB:K.series}). For $\zeta(\tau)$ as in (\ref{eB:Def.zeta}) we get for large enough $T$
\begin{align}\label{B.52}
\begin{split}
\Vert L^{(1,2)}(1-L^{(2,2)})^{-1}L^{(2,1)}\zeta\Vert &= \Big\Vert\sum^\infty_{r=0}K^{(r)}\zeta\Big\Vert\\
&\leq\sum^\infty_{r=0}\Vert K^{(r)}\zeta\Vert \leq \mcal O\klr{\frac1T}\Vert\zeta\Vert\ .
\end{split}
\end{align}
Here we use that for large enough $T$, $\sum^\infty_{r=0}\Vert K^{(r)}\zeta\Vert$ is bounded by a convergent geometric series due to (\ref{B.51}). The analogous argument was already used in (\ref{A.7}). 

In an analogous way we can estimate
\begin{align}\label{B.52a}
(1-L^{(2,2)})^{-1}L^{(2,1)}\zeta
=\sum^\infty_{r=0}(L^{(2,2)})^rL^{(2,1)}\zeta
\end{align}
where $\tau\rightarrow\zeta(\tau)$ is as in (\ref{eB:Def.zeta}). We find
\begin{align}\label{B.52b}
\Vert(L^{(2,2)})^rL^{(2,1)}\zeta\Vert
\leq\hat{C}_{21}(T\Delta\Gamma_{\rm min})^{-1}
\big[(N-M)C_{22}\Delta\Gamma^{-1}_{\rm min}\frac{2}{T}\big]^r\Vert\zeta\Vert\ ,
\end{align}
where 
\begin{align}\label{B.52c}
\hat{C}_{21} &= 2M C_{21}\max_{\substack{M+1\leq\alpha\leq N\\ \tau\in[0,\tau_0]}}
|\exp\left(\I\gamma_{\alpha\alpha}(\tau)\right)|\ .
\end{align}
With the argument of the geometric series we get, therefore, for large enough $T$
\begin{align}\label{B.52d}
\Vert(1-L^{(2,2)})^{-1}L^{(2,1)}\zeta\Vert
\leq{\cal O}\left(\frac{1}{T}\right)\Vert\zeta\Vert\ .
\end{align}

Consider next the operator $\tilde{L}$ (\ref{eB:Def.L}). From (\ref{B.52}) and (\ref{eB:norm.op.zeta}) we get, always for large enough $T$,
\begin{align}\label{B.53}
\begin{split}
\Vert\tilde{L}\zeta\Vert&=\Vert L^{(1,2)}(1-L^{(2,2)})^{-1}L^{(2,1)}
(1-L^{(1,1)})^{-1}\zeta\Vert\\
&\leq {\cal O}\left(\frac{1}{T}\right)\Vert(1-L^{(1,1)})^{-1}\zeta\Vert
\leq{\cal O}\left(\frac{1}{T}\right)\Vert\zeta\Vert\ .
\end{split}
\end{align}
Iterating this and using again the argument based on the geometric series leads to 
\begin{align}\label{B.54}
\Big\Vert\sum^\infty_{k=1}\tilde{L}^k\zeta\Big\Vert\leq{\cal O}\left(\frac{1}{T}\right)\Vert\zeta\Vert\ .
\end{align}

The next term to consider in (\ref{eB:chi.Solution}) is $(1-L^{(2,2)})^{-1}\tilde{\xi}^{(0)}$. We analyse this term in a way similar to $(1-L^{(1,1)})^{-1}$ above, see (\ref{eB:Def.zetac})-(\ref{eB:Tr.U.bound}). Let 
\begin{align}\label{B.55}
\eta_c=
\begin{pmatrix}
\eta_{c\,M+1}\\ \vdots\\ \eta_{c\,N}
\end{pmatrix}
\end{align}
be a constant vector and set as in (\ref{e4:tilde.xi.comp})
\begin{align}\label{B.56}
\begin{split}
\tilde{\eta}_\alpha(T;\tau) &= \exp\left[-\I\frac{T}{\tau_0}\varphi_\alpha(\tau)+\frac{T}{2\tau_0}\rho_1(\tau)
+\I\gamma_{\alpha\alpha}(\tau)\right]\eta_{c\alpha}\ ,\\
\alpha &\in \{M+1,\dots,N\}\ .
\end{split}
\end{align}
We define a matrix function $V(T;\tau)$ by
\begin{align}\label{B.57}
\sum^N_{\beta=M+1} V_{\alpha\beta}(T;\tau)\eta_{c\beta}
&= \klr{(1-L^{(2,2)})^{-1}\tilde{\eta}}_\alpha(T;\tau)\ ,\\
\begin{split}\label{B.58}
V(T;\tau) &= \Big(V_{\alpha\beta}(T;\tau)\Big)\ ,\\
\alpha,\beta &\in \{M+1,\dots,N\}\ .
\end{split}
\end{align}
This matrix satisfies
\begin{align}\label{B.59}
V(T;0) &= \mathbbm{1}_{N-M}\ ,\\ \label{B.60}
\frac{\partial}{\partial\tau}V(T;\tau) &= \tilde{A}(T;\tau)V(T;\tau)\ ,
\end{align}
where we get with (\ref{e2:Def.a})-(\ref{e2:Def.rho})
\begin{align}\label{B.61}
\begin{split}
\tilde{A}(T;\tau) &= \Big(\tilde{A}_{\alpha\beta}(T;\tau)\Big)\ ,\\
\tilde{A}_{\alpha\beta}(T;\tau) &= \left[-\I\frac{T}{\tau_0}E(\alpha,\tau)+\frac{T}{2\tau_0}\Gamma(1,\tau)\right]\delta_{\alpha\beta} + \I a_{\alpha\beta}(\tau)\ ,\\
\alpha,\beta &\in \{M+1,\dots,N\}\ .
\end{split}
\end{align}
In analogy to (\ref{eB:Def.B})ff. we consider
\begin{align}\label{B.62}
\tilde{B}(T;\tau)=V(T;\tau)V^\dagger(T;\tau)
\end{align}
and get
\begin{align}\label{B.63}
\begin{split}
\frac{\partial}{\partial\tau}\Tr\tilde{B}(T;\tau)
&= \Tr\klg{\tilde{B}(T;\tau)\kle{\tilde{A}(T;\tau)+\tilde{A}^\dagger(T;\tau)}}\\
&\leq\kle{-\frac{T}{\tau_0}\Delta\Gamma_{\rm min}+\frac{2c}{\tau_0}} \Tr\tilde{B}(T;\tau)\ ,
\end{split}
\end{align}
where we used (\ref{e4:LargerDecayRates}) and set 
\begin{align}\label{B.64}
c=\frac{\tau_0}{2}\max_{\tau\in[0,\tau_0]} 
\left\{\sum^N_{\alpha,\beta=M+1}
|a_{\alpha\beta}(\tau)-a^*_{\beta\alpha}(\tau)|^2\right\}^{1/2}\ .
\end{align}
From (\ref{B.63}) we get easily
\begin{align}\label{B.65}
\Tr\tilde{B}(T;\tau) &\leq (N-M)e^{2c}
\exp\left(-\frac{T}{\tau_0}\Delta\Gamma_{\rm min}\tau\right)\ ,\\ \label{B.66}
|V_{\alpha\beta}(T;\tau)| &\leq \sqrt{N-M}\ e^c
\exp\left(-\frac{T}{2\tau_0}\Delta\Gamma_{\rm min}\tau\right)
\end{align}
for all $\alpha,\beta\in\{M+1,\dots,N\}$. With the Cauchy-Schwarz inequality we get from (\ref{B.65}) also
\begin{align}\label{B.66a}
\begin{split}
\sum^N_{\alpha,\beta=M+1}|V_{\alpha\beta}(T;\tau)|
&\leq \kle{\klr{\sum_{\alpha,\beta=M+1}^N |V_{\alpha\beta}(T;\tau)|^2}\klr{\sum_{\alpha,\beta=M+1}^N 1^2}}^{1/2}\\
&= (N-M)\kle{\Tr\tilde{B}(T;\tau)}^{1/2}\\
&\leq(N-M)^{3/2}e^c
\exp\left(-\frac{T}{2\tau_0}\Delta\Gamma_{\rm min}\tau\right)\ .
\end{split}
\end{align}
From (\ref{B.66a}) we find
\begin{align}\label{B.66b}
\begin{split}
\left|\klr{(1-L^{(2,2)})^{-1}\tilde{\eta}}_\alpha(\tau)\right|
&= \left|\sum^N_{\beta=M+1}V_{\alpha\beta}(T;\tau)\eta_{c\beta}\right|\\
&\leq \hat{C}_{22}\exp\left(-\frac{T}{2\tau_0}\Delta\Gamma_{\rm min}\tau\right)\Vert\eta_c\Vert\ ,\\
\alpha &\in \{M+1,\dots,N\}\ ,
\end{split}
\end{align}
where
\begin{align}\label{B.66c}
\hat{C}_{22}=(N-M)^{3/2}e^c\ .
\end{align}

With the result (\ref{B.66a}) we can estimate the term 
\begin{align}\label{B.67}
L^{(1,2)}\big(1-L^{(2,2)}\big)^{-1}\tilde{\xi}^{(0)}
=L^{(1,2)}V\xi^{(0)}
\end{align}
in (\ref{eB:chi.Solution}), where $\xi^{(0)}$ is given in (\ref{e4:xi.initial}) and $\tilde{\xi}^{(0)}$ in (\ref{e4:tilde.xi.comp}). We get from (\ref{eB:L12}) 
\begin{align}\label{B.68}
\begin{split}
\klr{L^{(1,2)}V\xi^{(0)}}_\alpha(\tau) &= \I\int^\tau_0\mrmd\tau'\ 
\exp\kle{\I\frac{T}{\tau_0}\varphi_\alpha(\tau')-\frac{T}{2\tau_0}\rho_1(\tau')-\I\gamma_{\alpha\alpha}(\tau')}\\
&\quad\times \sum^N_{\beta,\gamma=M+1}a_{\alpha\beta}(\tau')V_{\beta\gamma}(T;\tau')\xi^{(0)}_\gamma\ ,\\
\alpha &\in \{1,\dots,M\}\ ;
\end{split}\\ \label{B.69}
\left|\klr{L^{(1,2)}V\xi^{(0)}}_\alpha(\tau)\right| &\leq
\frac{C'_{12}}{\tau_0}\Vert\xi^{(0)}\Vert\int^\tau_0\mrmd\tau'\ 
\sum^N_{\beta,\gamma=M+1}|V_{\beta\gamma}(T;\tau')|\ ,
\end{align}
where we set
\begin{align}\label{B.70}
C'_{12}=\max_{\substack{1\leq\alpha\leq M\\ M+1\leq\beta\leq N\\ \tau\in[0,\tau_0]}}
\tau_0|a_{\alpha\beta}(\tau)\exp\klr{-\I\gamma_{\alpha\alpha}(\tau)}|\ .
\end{align}
With (\ref{B.66a}) we obtain finally 
\begin{align}\label{B.71}
\Vert L^{(1,2)}(1-L^{(2,2)})^{-1}\tilde{\xi}^{(0)}\Vert
=\Vert L^{(1,2)}V\xi^{(0)}\Vert\leq\frac{2}{T}\Delta\Gamma^{-1}_{\rm min} C'_{12}(N-M)^{3/2}
e^c\Vert\xi^{(0)}\Vert\ .
\end{align}

Now we have collected all tools needed to discuss the solution for $\chi(T;\tau)$ given in (\ref{e4:chi.Solution}) and (\ref{eB:chi.Solution}) for large $T$. We have from (\ref{eB:chi.Solution})
\begin{align}\label{B.72}
\begin{split}
\chi &= \klr{1-L^{(1,1)}}^{-1}\kle{\chi^{(0)}+L^{(1,2)}\klr{1-L^{(2,2)}}^{-1}\tilde{\xi}^{(0)}}\\
&+ \klr{1-L^{(1,1)}}^{-1}\sum^\infty_{k=1}\tilde{L}^k
\kle{\chi^{(0)}+L^{(1,2)}\klr{1-L^{(2,2)}}^{-1}\tilde{\xi}^{(0)}}\ .
\end{split}
\end{align}
From (\ref{B.71}) we have 
\begin{align}\label{B.73}
\Vert L^{(1,2)}\klr{1-L^{(2,2)}}^{-1}\tilde{\xi}^{(0)}\Vert =
{\cal O}\left(\frac{1}{T}\right)\Vert\xi^{(0)}\Vert\ ,
\end{align}
With (\ref{B.54}) and (\ref{eB:norm.op.zeta}) we find
\begin{align}\label{B.74}
\Big\Vert\klr{1-L^{(1,1)}}^{-1}\sum^\infty_{k=1}\tilde{L}^k\chi^{(0)}\Big\Vert
&= {\cal O}\left(\frac{1}{T}\right)\Vert\chi^{(0)}\Vert\ ,\\ \label{B.75}
\Big\Vert\klr{1-L^{(1,1)}}^{-1}\sum^\infty_{k=1}\tilde{L}^k
L^{(1,2)}\klr{1-L^{(2,2)}}^{-1}\tilde{\xi}^{(0)}\Big\Vert
&= {\cal O}\left(\frac{1}{T^2}\right)\Vert\xi^{(0)}\Vert\ .
\end{align}
Inserting all this in (\ref{B.72}) gives 
\begin{align}\label{B.76}
\chi=\klr{1-L^{(1,1)}}^{-1}\chi^{(0)}
+{\cal O}\left(\frac{1}{T}\right)\Vert\chi^{(0)}\Vert
+{\cal O}\left(\frac{1}{T}\right)\Vert\xi^{(0)}\Vert\ .
\end{align}
With the definition of the matrix $\hat{U}(T;\tau)$, (\ref{eB:Def.U}), we get explicitly
\begin{align}\label{B.77}
\chi(T;\tau) = \hat{U}(T;\tau)\chi^{(0)}
+ {\cal O}\left(\frac{1}{T}\right)\Vert\chi^{(0)}\Vert
+ {\cal O}\left(\frac{1}{T}\right)\Vert\xi^{(0)}\Vert\ . 
\end{align}


\bibliography{myapvbib}

\end{document}